\documentclass[times,twocolumn,final]{elsarticle}
\usepackage{medima}
\usepackage{framed,multirow}
\usepackage{amsmath}
\usepackage{amssymb}
\usepackage{array}
\usepackage{latexsym}
\usepackage{url}
\usepackage{xcolor}
\usepackage{hyperref}
\hypersetup{pdfauthor={Name}}
\newcommand{\ie}{\textit{i}.\textit{e}.}

\usepackage{pifont}
\newcommand{\cmark}{\ding{51}}%
\newcommand{\xmark}{\ding{55}}%

\newcommand{\myparagraph}[1]{\vspace{0.1em}\noindent\textbf{#1}}

\usepackage{arydshln}
\usepackage{ulem}
\definecolor{newcolor}{rgb}{.8,.349,.1}
\newcommand{\vs}{-1mm}
\journal{Medical Image Analysis}
\begin{document}
\verso{Yi Lin, \textit{et~al.}}
\begin{frontmatter}
\title{Rethinking Boundary Detection in Deep Learning-Based Medical Image Segmentation}%
\author[1]{Yi \snm{Lin}\fnref{fn1}}
\author[2]{Dong \snm{Zhang}\fnref{fn1}}
\fntext[fn1]{These three authors contributed equally to this work.}
\author[1]{Xiao \snm{Fang}\fnref{fn1}}
\author[3]{Yufan \snm{Chen}}
\author[1,2]{Kwang-Ting \snm{Cheng}}
\author[1,3,4]{Hao \snm{Chen}\corref{cor1}}
\cortext[cor1]{Corresponding author: Hao Chen (E-mail:~jhc@cse.ust.hk).}
\address[1]{Department of Computer Science and Engineering, The Hong Kong University of Science and Technology, Hong Kong, China}
\address[2]{Department of Electronic and Computer Engineering, The Hong Kong University of Science and Technology, Hong Kong, China.}
\address[3]{Department of Chemical and Biological Engineering, The Hong Kong University of Science and Technology, Hong Kong, China}
\address[4]{HKUST Shenzhen-Hong Kong Collaborative Innovation Research Institute, Futian, Shenzhen, China. }
\begin{abstract}
Medical image segmentation is a pivotal task within the realms of medical image analysis and computer vision. 
While current methods have shown promise in accurately segmenting major regions of interest, the precise segmentation of boundary areas remains challenging. 
In this study, we propose a novel network architecture named CTO, which combines Convolutional Neural Networks (CNNs), Vision Transformer (ViT) models, and explicit edge detection operators to tackle this challenge. 
CTO surpasses existing methods in terms of segmentation accuracy and strikes a better balance between accuracy and efficiency, without the need for additional data inputs or label injections. 
Specifically, CTO adheres to the canonical encoder-decoder network paradigm, with a dual-stream encoder network comprising a mainstream CNN stream for capturing local features and an auxiliary StitchViT stream for integrating long-range dependencies. 
Furthermore, to enhance the model's ability to learn boundary areas, we introduce a boundary-guided decoder network that employs binary boundary masks generated by dedicated edge detection operators to provide explicit guidance during the decoding process. 
We validate the performance of CTO through extensive experiments conducted on seven challenging medical image segmentation datasets, namely ISIC 2016, PH2, ISIC 2018, CoNIC, LiTS17, and BTCV. 
Our experimental results unequivocally demonstrate that CTO achieves state-of-the-art accuracy on these datasets while maintaining competitive model complexity. 
The codes have been released at:~\href{https://github.com/xiaofang007/CTO}{CTO}.
\end{abstract}
\begin{keyword}
\KWD\\
Medical image segmentation\sep\\
Convolutional neural networks\sep\\
Vision Transformer\sep\\
Network architecture\sep\\
Boundary detection.
\end{keyword}
\end{frontmatter}
\section{Introduction}
\label{intro}
Medical image segmentation (MedISeg) is a fundamental yet challenging task that poses several significant research topics in both the medical image analysis and computer vision domains~\citep{zhang2022deep,lin2023rethinking,xiao2023transformers}. 
The primary objective of MedISeg is to accurately identify and localize semantic lesions or anatomical structures within medical images obtained through imaging modalities such as Computed Tomography, X-ray, and Magnetic Resonance Imaging~\citep{ronneberger2015u,lin2021seg4reg,zhang2022deep}. 
In recent years, MedISeg has garnered extensive attention and has been extensively investigated due to its potential applications in fields such as robotic surgery~\citep{gao2021future}, cancer diagnosis~\citep{lin2019automated,10528362}, and treatment planning~\citep{wijeratne2021learning}. 
To achieve satisfactory segmentation results in clinical practice, it is imperative to extract a comprehensive yet discriminative set of semantic feature representations~\citep{zhang2022deep,lin2023permutable}.
\begin{figure*}[t]
\centering
\includegraphics[width=1\textwidth]{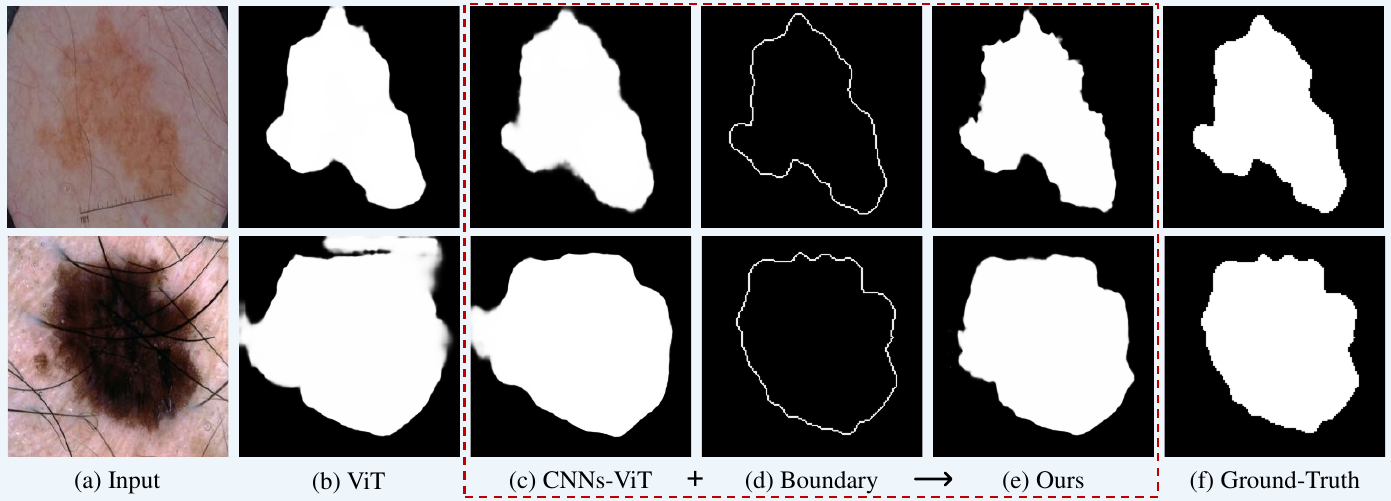}
\caption{Visualization comparisons of medical image segmentation results obtained from ViT~\citep{dosovitskiy2020image}, CNNs-ViT~\citep{hatamizadeh2022swin}, boundary detection operatorr~\citep{kanopoulos1988design}, and our proposed CTO. The results demonstrate that our method, which incorporates the explicit boundary detection, can achieve a significant improvement, especially in boundary areas, indicating the importance and effectiveness of the boundary detection operator in MedISeg. The red dashed bounding box highlights the core idea of our method. Samples are from the ISIC dataset~\citep{gutman2016skin,codella2019skin}.}
\label{fig:1}
\end{figure*} 
In the field of image processing, the vision transformer (ViT) architecture has emerged as a potential tool for enhancing accuracy in medical image recognition tasks when compared to methods based on convolutional neural networks (CNNs)~\citep{hatamizadeh2022unetr,zhang2022graph}. 
ViT-based approaches have achieved state-of-the-art performance in various fundamental medical image analysis tasks, such as image diagnosis~\citep{wu2022seatrans}, semantic segmentation~\citep{chen2021transunet}, and object detection~\citep{shamshad2022transformers}, primarily due to their ability to capture long-range feature dependencies. 
As illustrated in Figure~\ref{fig:1}(b), a ViT-based model effectively encompasses all the relevant pixels within the target area~\citep{dosovitskiy2020image}. 
A typical ViT-based MedISeg framework involves partitioning the input medical image into several image patches, which serve as image tokens for interactions through multi-head self-attention layers~\citep{vaswani2017attention}. 
Subsequently, a positional embedding layer may be employed to capture relative position information, if deemed necessary~\citep{zhang2023augmented}. 
Layer normalization strategies and feature regularization operations are then applied to generate the output predictions. 
These operations, along with specific layers, are interconnected to form a foundational transformer block, which is recurrently employed to encode valuable semantic representations for the task-specific head network.

Despite the success in capturing long-range feature dependencies, current ViT approaches exhibit limitations in terms of translation invariance and local feature representation~\citep{chen2021transunet,yuan2023effective}, which are crucial for MedISeg~\citep{zhang2022deep,shamshad2022transformers,yan2022after,xiao2023transformers}. 
To address these limitations, progressive methods, such as CNNs-ViT hybrid architectures, have been proposed for MedISeg, including TransUNet~\citep{chen2021transunet}, UNETR~\citep{hatamizadeh2022unetr}, and Swin UNETR~\citep{hatamizadeh2022swin}. 
These methods typically employ a specific network architecture as the primary framework and augment their learning capabilities by incorporating auxiliary learning modules~\citep{xiao2023transformers,shamshad2022transformers}.
For example, convolution layers are introduced to a ViT framework to capture local features, or Transformer layers are integrated into a CNNs framework to capture global features, enabling the ViT/CNNs model to acquire more informative yet elusive local/global features. 
Particularly, CNNs-ViT hybrid approaches for MedISeg are commonly based on the UNet framework~\citep{ronneberger2015u}, with multi-scale learning blocks and skip connections~\citep{huang2021missformer} incorporated into the backbone network~\citep{hatamizadeh2022unetr}. 
As depicted in Figure~\ref{fig:1}(c), the CNNs-ViT hybrid method achieves superior predictions in the local region of the foreground object when compared to the results in (b).

Boundary information matters in MedISeg~\citep{lin2023rethinking,wang2022boundary,lee2020structure}. 
Regrettably, current deep learning models often neglect this critical aspect in vision recognition tasks~\citep{wang2022boundary,lee2020structure,NEURIPS2020_07211688}. 
Explicitly incorporating boundary learning patterns, as opposed to implicit feature learning models, can offer significant advantages in terms of straightforward implementation, high efficiency, and purposeful objectives~\citep{fan2020camouflaged,wang2022boundary,bokhovkin2019boundary,wang2021boundary,hatamizadeh2019end}. 
Recently, edge detection operators have been successfully deployed in pixel-level recognition and generation tasks to explicitly enhance the learning capacity for localization~\citep{chen2021image,fan2020camouflaged,lin2022label}. 
In the context of MedISeg tasks, we firmly believe that the edge detection operator should assume a prominent role, as it can provide explicit priors for segmentation tasks. 
As depicted in Figure~\ref{fig:1}(d), the boundary of the foreground object-of-interest can be accurately extracted using the edge detection operator~\citep{kanopoulos1988design}. 
Moreover, the adoption of an explicit learning strategy can empirically enhance the feature representation capacity of the segmentation model.

In this paper, we introduce a novel network architecture, named CTO (\ie, Convolution, Transformer, and Operator), for MedISeg that seamlessly integrates CNNs, ViT, and edge detection operators to effectively harness local features, long-range feature dependencies, and explicit object boundaries. CTO follows the conventional encoder-decoder paradigm, wherein the encoder network comprises a CNNs network stream and a StitchViT network stream. 
The fusion of the feature maps obtained from these two streams serves as the input for the decoder network. 
To bolster the capacity for boundary learning, we introduce a boundary-guided decoder network that utilizes a self-generated boundary mask, extracted through edge detection operators, as explicit supervisory signals to guide the decoding learning process.
The proposed CTO can achieve higher segmentation accuracy and a better trade-off between accuracy and efficiency compared to current MedISeg methods. More importantly, CTO does not require any additional data inputs or label injections. 
As depicted in Figure~\ref{fig:1} (e), our method outperforms advanced CNNs-ViT hybrid architectures by effectively capturing all pixel areas of the object and generating precise predictions at the boundary regions. We thoroughly evaluate the performance of CTO on seven representative yet challenging MedISeg datasets, including ISIC 2016, ISIC 2018~\citep{gutman2016skin,codella2019skin}, PH2~\citep{mendoncca2013ph}, CoNIC~\citep{graham2021conic}, LiTS17~\citep{bilic2019liver}, MSD BraTS~\citep{antonelli2022medical}, and BTCV~\citep{irshad2022improved}. 
Our extensive experimental results demonstrate that CTO achieves the following: 1) higher accuracy on these datasets; 2) a considerable performance margin over state-of-the-art methods; 3) competitive model complexity and efficiency.

The main contributions of this paper are threefold: 
{
\begin{itemize} 
\item We investigate there fundamental components (\ie, CNN, Transformer, and Operator) by proposing a network architecture, CTO, that investigate the effectiveness of boundary detection in medical image segmentation from the perspective of feature extraction and information fusion.
\item We propose a novel StitchViT network that captures both global and local feature denpendencies while maintaining a low computational cost.
\item We propose a boundary-guided decoder network that uses a self-generated boundary mask as supervisions to enhance boundary learning capacity. 
\item Our proposed CTO achieves state-of-the-art accuracy on seven MedISeg benchmarks. Extensive ablation studies demonstrate its effectiveness and efficiency.
\end{itemize}
}
This paper significantly extends our previous work published at IPMI 2023~\citep{lin2023rethinking}. Specifically, (a) we enhance the network architecture of the ViT stream by proposing StitchViT network, which captures the global feature denpendencies while maintaining the local information in a low computational cost; (b) we provide a theoretical analysis to illustrate the rationale and effectiveness of our proposed method, and highlight its advantages in medical image segmentation; (c) we perform additional ablation studies, present more comprehensive experimental results, and provide more insightful result analyses.
\section{Related Work}
\begin{figure*}[t]
\centering
\includegraphics[width=.99\textwidth]{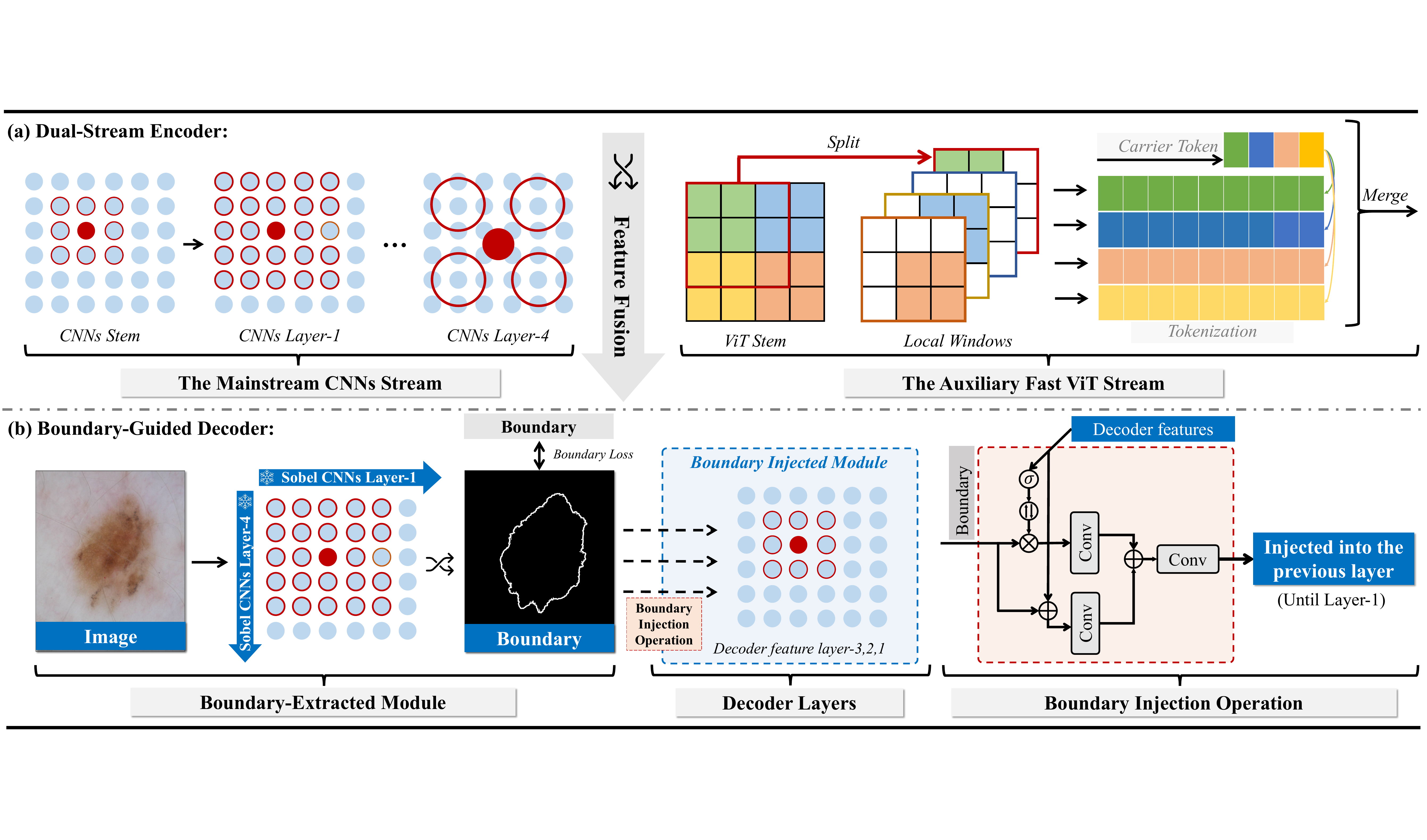}
\caption{(a) Illustration of our proposed Convolution, Transformer, and Operator (CTO), which follows an encoder-decoder paradigm. The dual-stream encoder network consists of a mainstream CNNs stream in the left side of (b) and an auxiliary ViT stream in the right side of (b), and the outputs of these two streams are fused together and used as the input of the decoder network. The boundary-guided decoder network in (c) employs an explicit boundary-enhanced module to guide its learning process. Specifically, the boundary detection operator is utilized to generate a self-generated boundary mask via the boundary-enhanced module that is then incorporated into the decoder network to enhance the boundary learning capacity and improve the segmentation accuracy. CTO integrates CNNs, ViT, and boundary detection into a unified framework. The red circle denotes the effective receptive field of the CNNs model, while the colored boxes represent the sliding windows of the ViT model.}
\vspace{\vs}
\label{fig:2}
\end{figure*}
\subsection{Medical Image Segmentation (MedISeg)}
Thanks to the rapid development of advanced image processing methods based on deep learning technology, MedISeg technology has also made significant progress and has been extensively studied~\citep{dosovitskiy2020image,xiao2023transformers,yu2022role}. MedISeg has also been applied in practical clinical applications. Existing methods can be broadly categorized into the following three categories: \romannumeral1) CNN-based methods, \romannumeral2) ViT-based methods, and \romannumeral3) CNN-ViT hybrid methods. 
\emph{In category \romannumeral1}, representative methods include VNet~\citep{milletari2016v}, UNet~\citep{ronneberger2015u}, Attention UNet~\citep{schlemper2019attention}, which uses CNNs as the backbone to extract fundamental image features and incorporate semantic segmentation tricks such as skip connections~\citep{zhang2021self}, multi-scale representation~\citep{chen2016dcan}, and feature interaction to enhance feature representations~\citep{chen2018voxresnet}. However, these methods may result in incomplete segmentation masks due to the inherently local nature of convolutions. 
\emph{In category \romannumeral2}, Swin-UNet~\citep{cao2021swin} and MissFormer~\citep{huang2021missformer} utilize ViT to aggregate long-range feature dependencies, replacing CNNs as the encoder/decoder. However, due to the limited number of annotated medical images and small inherent variability, these methods are difficult to optimize and have high computational costs. 
{\emph{In category \romannumeral3}, such as TransUNet~\citep{chen2021transunet}, UNETR~\citep{hatamizadeh2022unetr}, UNETR++~\citep{shaker2024unetr} and Swin-UNETR~\citep{hatamizadeh2022swin}, methods combine the advantages of both CNNs and ViT, capturing both local information and long-range feature dependencies~\citep{zhao2024semi}. However, these methods are computationally intensive and suffer from high computational overheads.} 
In this work, we propose a novel approach to leverage the benefits of both CNNs and ViT in the encoder process for improved MedISeg performance. We introduce StitchViT as a supplementary stream, which combines the advantages of both CNNs and ViT frameworks. Moreover, we incorporate an explicit edge detection operator to generate a self-generated boundary mask that guides the decoding learning process. 

\subsection{Operators in Image Processing}
In digital image processing domain, operators are mathematical functions that are used to extract relevant information, enhance images, or segment regions of interest~\citep{burger2022digital}. In general, edge detection operators in image processing tasks are a type of operator that detect edges or boundaries in digital images, which are important features in many computer vision tasks, including image segmentation. In image processing tasks, operators can mainly be divided into  two categories: first-order derivative operators and second-order derivative operators. For the \emph{first-order derivative operators}, there are Roberts, Prewitt, and Sobel~\citep{wang2007laplacian,kamgar1999optimally}. For the \emph{second-order derivative operators}, there is Laplacian~\citep{kanopoulos1988design}. These operators are commonly used for image edge detection tasks. Recently, edge detection operators have been applied to pixel-level computer vision tasks, such as manipulation detection and camouflaged object detection, where the goal is to detect subtle differences or manipulations in the image at a pixel-level~\citep{chen2021image,kanopoulos1988design,shi2023transformer}. In this work, we propose to use the Sobel operator as an explicit mask extractor to guide an implicit feature learning model for MedISeg. We utilize feature maps of the intermediate layer to synthesize a high-quality binary boundary prediction mask without requiring any additional data or annotations. 
\section{Methodology}
\label{sec:method}
\subsection{Overview}
\label{sec:arch}
{The overall network architecture of the proposed CTO is illustrated in Figure~\ref{fig:2}(a), which consists of a dual-stream encoder network in Figure~\ref{fig:2}(b) and a boundary-guided decoder network in Figure~\ref{fig:2}(c).} For an input medical image $X \in \mathbb{R}^{H\times W\times 3}$ with a spatial resolution of $H\times W$ and three channels, we aim to predict a pixel-wise label map (\ie, the semantic segmentation mask) $Y$, where each pixel has been assigned a predefined class label. The whole network is trained in an end-to-end manner following an encoder-decoder paradigm, which adopts skip connections to aggregate low-level features from the encoder network to the decoder network. 
Specifically, for the \emph{encoder network}, we propose a dual-stream encoder (\textit{ref.}~Sec.~\ref{sec:encoder}), which concurrently combines a mainstream CNNs network (\ie, Res2Net~\citep{gao2019res2net}) and an auxiliary lightweight vision transformer network with `Stitch' operation (\ie, StitchViT) to capture short-range feature dependencies and long-range feature dependencies, respectively. Such a combined architecture can address the problems of incomplete foreground (due to the lack of local features) and incomplete background (due to the lack of global features) of segmentation masks as much as possible. Particularly, thanks to the efficient sliding-window operation in StitchViT, such the combination will not bring many computational overheads. After fusing feature maps of CNNs and ViT streams via feature concatenation along the channel dimension, the resulting fused output is then utilized as input to \emph{decoder network}. 
In the boundary-guided \emph{decoder network} (\textit{ref.}Sec.\ref{sec:decoder}), an boundary-extracted module (BEM) is first employed, which utilizes a classical Sobel~\citep{kanopoulos1988design} edge detection operator to extract a binary boundary mask from the first and fourth layer features of the CNNs. Then, this self-generated boundary mask is incorporated into every intermediate layer of the \emph{decoder network} via the boundary-injected module (BIM) to guide its learning process, enabling the entire network to learn accurate object boundaries through deep supervision patterns~\citep{li2018deep}.
\subsection{Dual-Stream Encoder Network}
\label{sec:encoder}
\subsubsection{The CNNs Stream}
\begin{figure}[t]
\centering
\includegraphics[width=.25\textwidth]{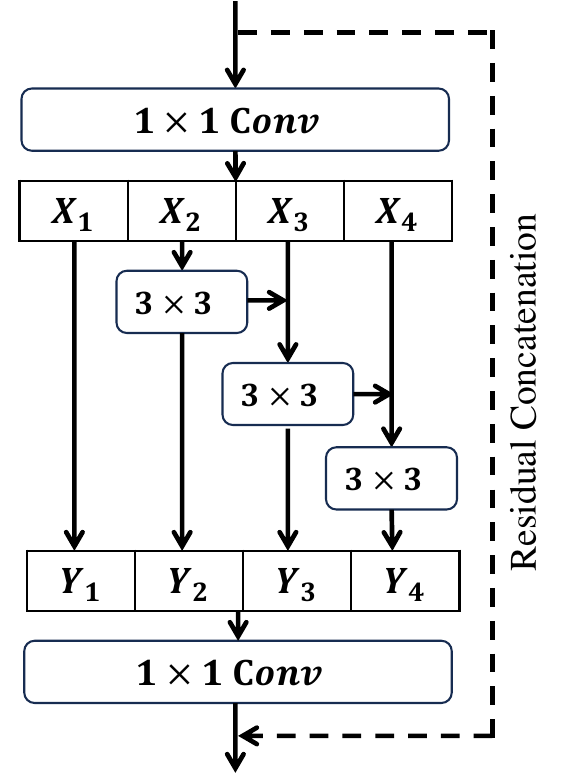}
\caption{Basic module of Res2Net~\cite{gao2019res2net}.}
\label{fig:cnn}
\end{figure}

The CNNs stream is utilized to capture short-range feature dependencies (\ie, local context features) of the input image. 
To accomplish this, we choose Res2Net~\citep{gao2019res2net}, an efficient and powerful CNNs backbone that comprises one convolutional stem and four Res2Net blocks. Each Res2Net block consists of several basic Res2Net modules, where each module has a scale dimension of 4. 
As illustrated in Figure~\ref{fig:cnn}, the basic Res2Net module is a variant of the classical ResNet module~\citep{he2016deep} that employs the split attention mechanism to capture multi-scale feature representations, \ie, $X_1$ to $X_4$. 
These multi-scale features obtained in each module are first concatenated together in a cascaded manner via $3 \times 3$ convolutional operation (\ie, $Y_1$ to $Y_4$) and then concatenated along the channel dimension using a $1 \times 1$ convolutional operation to form the output of the basic Res2Net module.
Based on Res2Net, the CNNs backbone network can generate feature maps $F_c$ with a spatial resolution of $H/4 \times W/4$, $H/8 \times W/8$, $H/16 \times W/16$, and $H/32 \times W/32$, respectively.  $c = 1,2,...,4$ denotes the index of the layer where the output feature is located. Unless stated otherwise, we follow the default architecture and settings of the Res2Net backbone as described in the referenced paper.
\subsubsection{The Auxiliary ViT Stream}
\label{sec:mst}
\begin{figure}[t]
    \centering
    \includegraphics[width=0.48\textwidth]{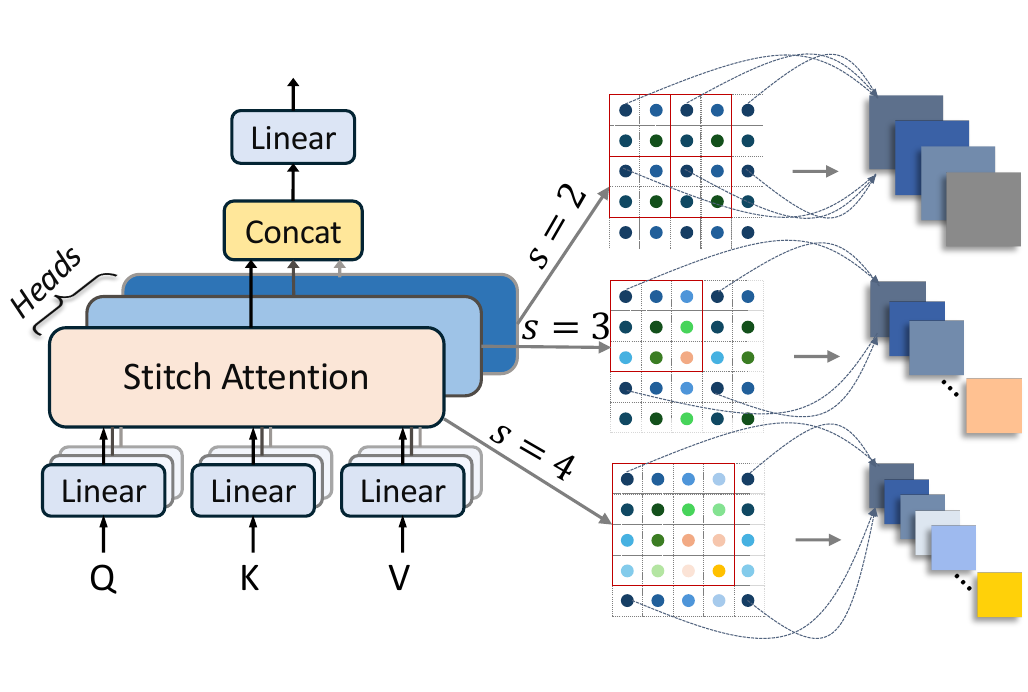}
    \caption{Illustration of Stitch-Vit. First, the channels of the feature map are divided into different heads. Then, ``stitch'' operation is performed to sample the feature map using different stitch rates in different heads. Finally, the sampled feature maps are concatenated and transformed to the original feature map. The red box highlights the stitch rates of our method, and the blue points reveal the sampling positions.}
    \label{fig:stitchtrans}
    \end{figure}

In addition to short-range feature dependencies, long-range feature dependencies also matter for MedISeg~\citep{chen2021transunet,zhang2022deep,yuan2023effective,hatamizadeh2023fastervit}. 
To address this, we introduce the StitchViT, a vision transformer within the CTO framework. 
The StitchViT effectively captures long-range feature dependencies across different feature scales through the stitch operation.
Figure~\ref{fig:stitchtrans} illustrates the implementation process of a StitchViT block on the input feature map $\text{F}_c \in \mathbb{R}^{\frac{H}{4}\times \frac{W}{4}\times C}$, where $C$ is the number of channel. 
Initially, the input feature map is divided into multiple patches using the stitch operation. 
We define a dilation rate $s \in \mathbb{N}^+$ to control the degree of sparsity.
In this study, we set the stitch rate to $s=2,4,8,16$ to obtain multi-scale feature patches with diverse receptive fields.

For instance, when considering a stitch rate of $s=2$, the corresponding patches are obtained by sampling the feature map with a stride of $2$. 
This process can be defined as follows:
\begin{equation}
    \begin{aligned}
        \text{P}_1 &= \text{F}_c[1::2, 1::2, :], \\
        \text{P}_2 &= \text{F}_c[1::2, 2::2, :], \\
        \text{P}_3 &= \text{F}_c[2::2, 1::2, :], \\
        \text{P}_4 &= \text{F}_c[2::2, 2::2, :],
    \end{aligned}
\end{equation}
where $\text{P}_i$ denotes the $i$-th patch, and $[a::b]$ denotes the sampling operation with a stride of $b$ starting from $a$.
Subsequently, self-attention is performed on each patch to capture the long-range feature dependencies:
\begin{equation}
\text{A} = \text{softmax}\left(\frac{QK^T}{\sqrt{d_k}}\right)V,
\end{equation}
where $Q$, $K$, and $V$ are the query, key, and value matrices, respectively, which are obtained by linearly projecting the input patch embeddings as $Q=W_q\text{P}_i$. $d_k$ is the dimension of the key.
Empirically, truncating the feature vectors within a certain range does not diminish the model's recognition performance. Instead, it can significantly reduce the computational cost~\citep{zhang2020feature}. The resulting feature representation from the MHSA layer is then passed through the feed-forward network (FFN) to obtain the output $\text{F}_t$:
\begin{equation}
    \text{F}_t = \text{FFN}(\text{A}),
\end{equation}
where $\text{FFN}$ consists of two $3\times3$ convolutional layers with the ReLU activation function. 
Finally, $\text{F}_t$ is reshaped into the same size as $\text{F}_c^4$ to obtain the output. 
The remaining transformer blocks are processed similarly. 
All outputs of the transformer blocks are concatenated along the channel dimension and fed into a $3\times3$ convolutional layer to obtain the final output.

{StitchViT ingeniously orchestrates the assimilation of long-range dependencies by seamlessly stitching together localized features from multiple overlapping patches. Furthermore, this innovative methodology excels in harmonizing the aggregation of global insights while preserving the minutiae of local details, a critical facet in the nuanced landscape of medical image segmentation endeavors.}

\subsection{Boundary-Guided Decoder Network}
\label{sec:decoder}
The boundary-guided decoder network incorporates a boundary-extracted module to extract the explicit boundary information of foreground objects. 
The boundary-enhanced image feature $F_b$ is then integrated into the multi-level encoder's features by a boundary-injected module. 
This aims to simultaneously characterize the intra- and inter-class consistency in the feature space, enriching the feature representative ability.

\subsubsection{Boundary-Extracted Module (BEM)}
BEM takes the high-level $F_c^4$ and low-level features $F_c^1$ as inputs to extract the boundary mask while filtering out the trivial boundary-irrelevant information. 
To achieve this, we employ the widely used Sobel operator~\citep{kanopoulos1988design} in both the horizontal ($G_x$) and vertical ($G_y$) directions to obtain gradient maps. 
Specifically, we utilize two $3\times 3$ parameter-fixed operators with a stride of 1. 
The definitions of these operators are as follows:
\begin{equation}
    K_x = \begin{bmatrix}
        -1 & 0 & 1 \\
        -2 & 0 & 2 \\
        -1 & 0 & 1
    \end{bmatrix}, \quad
    K_y = \begin{bmatrix}
        -1 & -2 & -1 \\
        0 & 0 & 0 \\
        1 & 2 & 1
    \end{bmatrix}.
\end{equation}
Next, we apply these operators to the input feature map to obtain the gradient maps $M_x$ and $M_y$. 
The gradient maps are then normalized using a sigmoid function and fused with the input feature map to yield the boundary-enhanced feature map $F_e$:
\begin{equation}
    F_e = F_c \odot \sigma(M_{xy}),
\end{equation}
where $\odot$ denotes the element-wise multiplication operation, $\sigma$ is the sigmoid activation function, and $M_{xy}$ is the concatenation operation of $M_x$ and $M_y$ along the channel dimension.
We proceed to fuse the edge-enhanced feature maps $F_e^1$ and $F_e^4$ using a simple stacked convolution layer in the bottleneck. 
{
Specifically, we first apply a $1\times 1$ convolution followed by bilinear upsampling to match the size of $F_e^1$. 
Then, we separately apply a $1\times 1$ convolution layer to align the channel sizes of these two features. Finally, we concatenate the resulting feature maps along the channel dimension and apply a two-layer $3\times 3$ convolution to obtain the final feature map $\bar{F}_e$.
The output is supervised by the ground truth boundary map, which eliminates the noise in the edge features, producing the boundary-enhanced feature $F_b$.
}
\subsubsection{Boundary-Injected Module (BIM)}
The enhanced features generated by BEM can serve as a prior to enhance the image segmentation capability. 
In this regard, we propose the Boundary-injected module (BIM), which integrates the boundary-enhanced feature $F_b$ into the decoder network.
At the core of BIM lies the Boundary Injection Operation (BIO), which employs a dual-path boundary fusion scheme to enhance the feature representation in both the foreground and background. 
BIO takes two inputs: the channel-wise concatenation of the boundary-enhanced feature $F_b$ and the corresponding feature $F_c$ from the encoder network, as well as the feature from the previous decoder layer $F_d^{j-1}$.
{These two inputs are fed into BIO, as depicted in Figure~\ref{fig:2}(c)}, which consists of two separate paths designed to enhance the feature representation in the foreground and background, respectively. 
For the foreground path, we concatenate the two inputs along the channel dimension and apply sequential \textit{Conv-BN-ReLU} (\ie, $3\times3$ convolution, batch normalization, ReLU) layers to obtain the foreground feature $F_{fg}$. 
As for the background path, we incorporate a background attention component to selectively focus on the background information. This is expressed as:
\begin{equation}
    F_{bg} = \text{Convs}\left ((1-\sigma (F_d^{j-1})) \odot F_c \right ),
\end{equation}
where $\text{Convs}$ is a three-layer \textit{Conv-BN-ReLU} layer, $\sigma$ is the sigmoid function, and $\odot$ denotes the element-wise multiplication.
The term $\left(1-(\sigma F_d^{j-1})\right)$ is the background attention map, which is computed by first applying the sigmoid function to the feature map from the previous decoder layer, generating a foreground attention map. Then, we subtract the foreground attention map to obtain the background attention map. Finally, we concatenate the foreground feature $F_{fg}$, the background feature $F_{bg}$, and the previous decoder feature $F_d^{j-1}$ along the channel dimension to obtain the final output $F_d^j$.
\subsection{Overall Loss Function}
\label{sec:allloss}
Intrinsically, CTO is a multi-task learning model that performs both semantic segmentation and boundary detection in an uniform recognition framework. We define the overall loss function to jointly optimize these two tasks.

\subsubsection{Segmentation Loss.}
Following MedISeg model is commonly used, the semantic segmentation loss is the weighted sum of cross-entropy loss $\mathcal{L}_{\text{CE}}$ and mean intersection-over-union (mIoU) loss $\mathcal{L}_{\text{mIoU}}$. They are defined as follows:
\begin{equation}
    \mathcal{L}_{\text{CE}} = -\frac{1}{N}\sum_{i=1}^N \left(y_i \log(\hat{y}_i) + (1-y_i) \log(1-\hat{y}_i) \right),
\end{equation}
\begin{equation}
    \mathcal{L}_{\text{mIoU}} = 1 - \frac{\sum_{i=1}^N (y_i * \hat{y}_i)}{\sum_{i=1}^N (y_i + \hat{y}_i - y_i * \hat{y}_i)},
\end{equation}
where $y_i$ and $\hat{y}_i$ are the ground truth and predicted label for the $i$-th pixel, respectively, and $N$ is the total number of pixels in the image.

\subsubsection{Boundary Loss.}
To avoid the occurrence of class imbalance between foreground and background pixels in boundary detection, we also employ the Dice Loss:
\begin{equation}
    \mathcal{L}_{\text{Dice}} = 1 - \frac{2 \sum_{i=1}^N (y_i * \hat{y}_i)}{\sum_{i=1}^N (y_i + \hat{y}_i)}.
\end{equation}

\subsubsection{Total Loss.}
The total loss is composed of the major segmentation loss $\mathcal{L}_{\text{seg}}$ and boundary loss $\mathcal{L}_{\text{bnd}}$. For the boundary detection loss, we only consider the prediction from BEM, which takes encoder's feature maps from the high-level layer (\ie, $F_b^4$) and low-level layer (\ie, $F_b^1$) as input. As for the major image segmentation loss, we apply the deep supervision strategy to obtain the prediction from the decoder's feature at different levels. The total loss is formulated as:
\begin{equation}
    \mathcal{L} = \mathcal{L}_{\text{seg}} + \mathcal{L}_{\text{bnd}}
                = \sum_i^L \left(\mathcal{L}_{\text{CE}} + \mathcal{L}_{\text{mIoU}}\right)  + \alpha\mathcal{L}_{\text{Dice}},
\end{equation}
where $L$ is the number of BIM, which is set to 3 in this work. $\alpha$ is the weighting factor, which is set to 3.

\subsection{Discussion}
\label{theoretical}
{In this section, we discuss the underlying motivation behind the proposed CTO. 
Based on the minimum redundancy maximum relevance (MRMR) principle, the feature set that is mutually and maximally dissimilar and relevant to the target holds superior feature representation ability~\citep{peng2005feature} and generalizability.
Formally, given a feature set $\mathcal{F} = {F_1, F_2, ..., F_n}$, the MRMR principle aims to find a subset $\mathcal{S} \subset \mathcal{F}$ that maximizes the mutual information $I(\mathcal{S}; Y)$ with a response variable $Y$, while minimizing the redundancy $I(\mathcal{S}; \mathcal{F})$:
\begin{equation}
    \text{MRMR}(\mathcal{F}) = \mathop{\text{argmax}}\limits_{\mathcal{S} \subset \mathcal{F}} \underbrace{\frac{1}{|\mathcal{S}|}\sum_{f\in\mathcal{S}}I(F,Y)}_{\text{Relevance}} - \underbrace{\frac{1}{|\mathcal{S}|^2}\sum_{f\in\mathcal{S}}I(F,\mathcal{F})}_{\text{Redundancy}}.
\end{equation}
}
In the context of CTO, our main objective is to leverage the strengths of CNN, Transformer, and the edge detection operator to achieve accurate medical image segmentation. 
Specifically, the CNN stream within the encoder network is designed to capture local information, while the Transformer stream focuses on capturing global information. 
Additionally, the edge detection operator is utilized to extract crucial boundary information of the object.
The features extracted from these three modules are highly relevant to the target in the medical image segmentation task, while simultaneously exhibiting dissimilarity with each other. 
For instance, in the case of abdominal organ segmentation using the BTCV dataset, the CNN stream is adept at capturing contextual information pertaining to the organ surface, the Transformer stream excels at capturing global semantic information such as position, and the edge detection operator effectively highlights the contrast between the organ and its background.
By combining these three modules, we can effectively enhance the feature representation capabilities of the model and improve its generalization ability, thereby facilitating accurate medical image segmentation.
\section{Experiments}
\label{sec:exp}
\subsection{Datasets and Evaluation Metrics}
\label{sec:dataset}
\myparagraph{Datasets.} In our experiments, we evaluate the performance of CTO on seven publicly available MedISeg datasets, namely ISIC 2016~\citep{codella2019skin}, ISIC 2018~\citep{codella2019skin}, PH2~\citep{mendoncca2013ph}, the Colon Nuclei Identification and Counting (CoNIC) challenge dataset~\citep{graham2021conic}, the Liver Tumor Segmentation (LiTS17) challenge dataset~\citep{bilic2019liver}, MSD BraTS~\citep{antonelli2022medical}, and the Beyond the Cranial Vault (BTCV) challenge dataset~\citep{vaswani2017attention}. 
The brief introduction of these experimental datasets is as follows: 
\begin{itemize}
\item ISIC 2016~\citep{codella2019skin} and ISIC 2018~\citep{codella2019skin} are datasets of dermoscopic lesion images for melanoma segmentation, these two datasets are available in the International Skin Imaging Collaboration (ISIC) challenge. ISIC 2016 contains 1187 images, while ISIC 2018 contains 2594 images. 
\item PH2~\citep{mendoncca2013ph} is a dataset of dermoscopic images for melanoma detection and classification, which consists of 200 images.
\item The CoNIC challenge dataset~\citep{graham2021conic} is a collection of colon tissue images for nuclei segmentation and counting, which contains 4,981 non-overlapping image patches of size $256\times256$.
\item The LiTS17 challenge dataset~\citep{bilic2019liver} is a collection of abdominal CT scans for liver tumor segmentation, which contains 130 training and 70 testing cases.
\item MSD BraTS~\citep{antonelli2022medical} consists of 484 MRI scans of brain tumor, with four modalities: FLAIR, T1w, T1gd, and T2w. 
\item The BTCV dataset~\citep{vaswani2017attention} is a collection of abdominal CT scans for organ segmentation. We use the 30 abdominal CT scans in the MICCAI 2015 Multi-Atlas Abdomen Labeling Challenge, with 3779 axial contrast-enhanced abdominal clinical CT images in total.
\end{itemize} 

{On the datasets of ISIC 2018, CoNIC, and LiTS17 datasets, we perform 5-fold cross-validation.
We use ISIC 2016 as our training dataset and employ PH2 as our independent test set to assess the generalizability of the model.
MSD BraTS is divided into 80\%, 15\%, and 5\% for training, validation, and test.
For BTCV, we follow the approach of previous studies~\citep{cao2021swin,chen2021transunet} and split the dataset into 18 cases for training and 12 cases for testing.}

\myparagraph{Evaluation Metrics.} We employ commonly used evaluation metrics in MedISeg, including Dice Coefficient (Dice), Intersection over Union (IoU), average Hausdorff Distance (HD), and Panoptic Quality (PQ)~\citep{cao2021swin,chen2021transunet,lee2020structure}. Additionally, we evaluate the model efficiency using the number of Floating-point Operations (FLOPs) and model Parameters (Params.). These metrics provide a comprehensive evaluation of the segmentation performance and computational efficiency of our proposed method.

\subsection{Baseline Models}
\label{sec:baselines}

\subsection{Implementation Details}
\label{sec:implementation}
The proposed CTO is optimized using the Adam optimizer with an initial learning rate of $1\mathrm{e}{-4}$. 
The batch size is set to 32, and the input image size was $256\times256$. The encoder network is initialized with the pre-trained weights of Res2Net-50~\citep{gao2019res2net} on ImageNet~\citep{deng2009imagenet} and then fine-tuned for 90 epochs on a single NVIDIA RTX 3090 GPU. All 3D volumes are inferred in a sliding-window manner with a stride of 1 in the $z$-axis direction, and the final segmentation results are obtained by stacking the prediction maps to reconstruct the 3D volume for evaluation. Except for a special statement, all experimental settings, including data pre-processing procedures, follows the previous paper~\citep{cao2021swin,lee2020structure}. The experimental design is carefully planned to ensure that the proposed CTO is optimized and evaluated under rigorous conditions.

\subsection{Ablation Study}
Our ablation study aims to explore the effectiveness of each component (\ie, dual-stream encoder, CBM, BEM, and BIM) in CTO and demonstrate the superiority of boundary-guided decoder. To achieve these goals, we add different components to the baseline model, and compared boundary-guided decoder with the state-of-the-art (SOTA) boundary detection methods currently used for MedISeg.

\begin{table}[!t]
    \small
    \begin{center}
    \caption{Result comparisons (\%) on ISIC 2018 for selecting of the encoder network. ``\emph{w/}'' denotes with the corresponding module implementation.}
    \renewcommand\arraystretch{1.2}
    \resizebox{0.49\textwidth}{!}{
    \setlength{\tabcolsep}{3pt}
    {
    \begin{tabular}{ l c  c  | c c c c c} 
    \hline \hline
    Methods &   &   &    Dice$\uparrow$ & IoU$\uparrow$ & GFLOPs$\downarrow$ & Params$\downarrow$ & Memory$\downarrow$ \\ 
    \hline
    CNNs-ViT & & & & & & \\
    \cdashline{1-8}[0.8pt/2pt]
    TransUNet &  &  &   89.97 & 83.30 & 37.02 & 114.65 & 7302 \\
    TransFuse &  &  &   89.62  & 82.89 & 63.84 & 71.27 & 6914 \\
    MedFormer  &  &  &   89.15 & 82.19 & 21.02 & 80.44 & 5564 \\
    LeViT &  &  &   86.98  & 79.66 & 23.57 & 64.23 & 6734 \\
    Sequential (Res2Net + StitchVit) & &  & 
     89.86 & 83.02 & 20.67 & 154.25 & 5492 \\
    Parallel (Res2Net + StitchVit) &  &  &   \textbf{90.63} & \textbf{83.97} & 22.70 & 62.22 & 4104 \\
    \hline
    CNNs-based (+StitchVit) & & & & & & \\
    \cdashline{1-8}[0.8pt/2pt]
    ResNet &  &  &    90.28 & 83.63 & 31.10 & 61.66 & 3984  \\
    ResNeXt &  &  &   89.53 & 82.76 & 22.35 & 61.13 & 4020  \\
    Res2NeXt &  &  &    90.42 & 83.82 & 22.28 & 60.78 & 4098  \\
    Res2Net &  &  &   \textbf{90.63} & \textbf{83.97} & 22.70 & 62.22 & 4104 \\
    \hline
    Transformer-based (+Res2Net) & & & & & & \\
    \cdashline{1-8}[0.8pt/2pt]
    ViT &  &  &   90.38 & 83.79 & 22.70 & 61.82 & 4022 \\
    SwinV2 &  &  &   90.30 & 83.77 & 23.13 & 62.04 & 7458 \\
    FasterViT &  &    & 90.22 & 83.62 & 22.83 & 62.32 & 5568 \\
    Twins &  &    & 90.18 & 83.59 & 23.47 & 62.37 & 9778 \\
    StitchVit &  &    & \textbf{90.63} & \textbf{83.97} & 22.70 & 62.22 & 4104 \\
    \hline \hline
    \end{tabular}
    \label{tab1}
    }
    }
    \end{center}
\end{table}
\myparagraph{Effectiveness of Dual-Stream Encoder.}
In this section, we conduct an experimental analysis to assess the feature extraction capabilities of different encoders. 
The results are presented in Table~\ref{tab1}.
First, on the ISIC 2018 dataset, we compare our dual-stream encoder with SOTA hybrid encoders that combine CNNs and Transformers, including TransUNet~\citep{chen2021transunet}, TransFuse~\citep{zhang2021transfuse}, MedFormer~\citep{gao2022data}, LeViT~\citep{graham2021levit}.
Notably, our dual-stream encoder achieves the highest performance with a Dice score of 90.63\% and an IoU score of 83.97\%, surpassing the performance of the other methods.

Furthermore, in order to evaluate the individual effectiveness of the CNNs and Transformers within the dual-stream encoder, we conduct additional experiments by substituting Res2Net~\citep{gao2019res2net} with ResNet~\citep{he2016deep}, ResNeXt~\citep{xie2017aggregated}, and Res2NeXt~\citep{gao2019res2net}, and replacing StitchViT with ViT~\citep{dosovitskiy2020image}, SwinV2~\citep{liu2022swin}, FasterViT~\citep{hatamizadeh2023fastervit}, and Twins~\citep{chu2021twins}.
{We also compare our dual-stream encoder with the sequential encoder, where the CNNs and Transformers are concatenated in a sequential manner.
The results presented in Table~\ref{tab1} demonstrate the consistent superiority of our dual-stream encoder compared to other methods. This reaffirms the effectiveness of our dual-stream encoder in capturing both local and global information, thereby enhancing its performance in the context of MedISeg.}

\begin{table}[t]
\small
\begin{center}
\renewcommand\arraystretch{1.2}
\setlength{\tabcolsep}{1pt}{
\caption{Result comparisons (\%) on ISIC 2016~\citep{codella2019skin} for validating the superiority of our boundary enhanced module. ``\emph{w/}'' denotes with the corresponding module implementation.} 
\begin{tabular}{ l c  c  c |
>{\centering\arraybackslash}p{1cm} 
>{\centering\arraybackslash}p{1cm} 
>{\centering\arraybackslash}p{1cm} 
}
\hline \hline
Settings &   &   &   & Dice~$\uparrow$ & IoU~$\uparrow$ & \scriptsize{GFLOPs$\downarrow$} \\ 
\hline
Baseline & & & & 91.59 & 84.92 & 17.13 \\
\cdashline{1-7}[0.8pt/2pt]
\emph{w/} BATR~\citep{wang2021boundary} & & & & 91.62 & 85.03 & 45.99 \\
\emph{w/} XBound-Former~\citep{wang2023xbound} & & &  & 92.11 & 86.07 & 20.84 \\
\emph{w/} PEE~\citep{wang2022boundary} & & & & 91.12 & 84.65 & 23.45 \\
\emph{w/} Boundary Loss~\citep{kervadec2019boundary} & & & & 91.75 & 85.25 & 17.13 \\
\emph{w/} Boundary DoU Loss~\citep{sun2023boundary} & & & & 92.01 & 85.70 & 17.13 \\
\emph{w/} BEM (ours) & & & & \textbf{92.56} & \textbf{86.60} & 22.70 \\
\hline \hline
\end{tabular}
\label{tab2}}
\end{center}
\vspace{-4mm}
\end{table}
\begin{table}[!t]
\small
\caption{Ablation study results (\%) on ISIC 2018~\citep{codella2019skin}. 
We choose Res2Net~\citep{gao2019res2net} and a lightweight ViT~\citep{dosovitskiy2020image} as the baseline models. ``*'' means the component achieves significant performance improvement with p < 0.05 via paired t-test.}
\label{tab3}
\centering
\renewcommand\arraystretch{1.2}
\setlength{\tabcolsep}{4.5pt}{
\begin{tabular}{ c c c c c | c c} 
\hline \hline
 CNNs & Stitch-ViT & CBM & BEM & BIM & Dice~$\uparrow$ & IoU~$\uparrow$ \\
\hline
\cmark & \xmark & \xmark & \xmark & \xmark & 89.18 & 82.46 \\  
\xmark & \cmark & \xmark & \xmark & \xmark & 87.53 & 79.89 \\  
\cmark  & \cmark & \xmark & \xmark & \xmark & 89.60$^*_{\color{red}{+0.42}}$ & 82.92$^*_{\color{red}{+0.46}}$ \\
\cmark & \cmark & \cmark & \xmark & \xmark & 89.74$^*_{\color{red}{+0.56}}$ & 83.03$_{\color{red}{+0.57}}$ \\
\cmark & \cmark & \cmark & \cmark & \xmark & 90.19$_{\color{red}{+1.01}}$ & 83.56$_{\color{red}{+1.10}}$ \\
\cmark & \cmark & \cmark & \cmark & \cmark & 90.63$^*_{\color{red}{+1.45}}$ & 83.97$^*_{\color{red}{+1.51}}$ \\
\hline \hline
\end{tabular}}
\end{table}
\begin{table*}[!t]
\small
\caption{Result comparisons (\%) with the state-of-the-art methods on ISIC~\citep{gutman2016skin,codella2019skin} \& PH2~\citep{mendoncca2013ph}. ``Nan'' denotes that this model is based on shallow learning and does not have a backbone.}
\label{tab4}
\centering
\renewcommand\arraystretch{1.2}
\setlength{\tabcolsep}{7pt}{
\begin{tabular}{ r c c c | r c c c} 
\hline \hline
\multirow{2}{*}{Methods} & \multirow{2}{*}{Backbone} & \multicolumn{2}{c|}{ISIC 2016 \& PH2} & \multirow{2}{*}{Methods} & \multirow{2}{*}{Backbone} & \multicolumn{2}{c}{ISIC 2018}\\
 & & Dice~$\uparrow$ & IoU~$\uparrow$ & & & Dice~$\uparrow$ & IoU~$\uparrow$ \\
\hline
SSLS~\citep{ahn2015automated} & Nan & 78.38 & 68.16 & Deeplabv3~\citep{chen2017rethinking} & ResNet-50 & 88.4& 80.6\\
MSCA~\citep{bi2016automated} & Nan  & 81.57 & 72.33 & U-Net++~\citep{zhou2018unetpp} & VGG-16 & 87.9& 80.5\\
FCN~\citep{long2015fully} & VGG-16  & 89.40 & 82.15 & CE-Net~\citep{gu2019cenet} & ResNet-34        & 89.1& 81.6\\
MFCN~\citep{bi2017dermoscopic} & VGG-16  & 90.66 & 83.99 & MedT~\citep{valanarasu2021medical} & ViT-B16 & 85.9 & 77.8\\
SBPSeg~\citep{lee2020structure} & VGG-16  & \underline{91.84} & \underline{84.30} &TransUNet~\citep{chen2021transunet} & ResNet-50  & \underline{89.4}& \underline{82.2}\\
\cdashline{1-8}[1.5pt/1.5pt]
MedSAM~\citep{ma2023segment} &ViT-B16 & 84.67 & 78.20 &MedSAM~\citep{ma2023segment} &ViT-B16   & 83.5 & 77.4\\
\cdashline{1-8}[1.5pt/1.5pt]
CTO(Ours) & CNNs-ViT  & \textbf{92.56} & \textbf{86.60} & Ours & CNNs-ViT  & \textbf{90.6} & \textbf{84.0} \\
\hline \hline
\end{tabular}}
\vspace{-4mm}
\end{table*}
\begin{table*}[!t]
\small
\caption{Result comparisons (\%) with the state-of-the-art methods on CoNIC~\citep{graham2021conic} and LiTS17~\citep{bilic2019liver}.}
\label{tab5}
\centering
\renewcommand\arraystretch{1.2}
\setlength{\tabcolsep}{3pt}{
\begin{tabular}{ r | c | c c c | c c | c c c} 
\hline \hline
\multirow{2}{*}{Methods} & \multirow{2}{*}{Backbone} & \multicolumn{3}{c|}{CoNIC} & \multicolumn{2}{c|}{LiTS17} & \multicolumn{3}{c}{Complexity}\\
 & & Dice~$\uparrow$ & IoU~$\uparrow$ & PQ~$\uparrow$  & Dice~$\uparrow$ & IoU~$\uparrow$ & \scriptsize{Params.(M)} & \scriptsize{GFLOPs} & \scriptsize{Memory(MB)}\\
\hline
V-Net~\citep{milletari2016v} & ResNet-50 & 78.39$_{3.397}$ & 65.15$_{4.287}$ & 62.12$_{7.660}$ & 82.63$_{0.420}$  & 70.79$_{0.669}$ & 11.84 & 18.54 & 3452\\ 
U-Net~\citep{ronneberger2015u} & VGG-16 & 79.70$_{1.388}$ & 67.25$_{1.798}$ & 66.09$_{2.299}$ & 79.50$_{2.111}$   & 66.64$_{2.738}$ & 7.78  & 14.59 & 3506\\  
R50-UNet~\citep{ronneberger2015u} & ResNet-50 &  78.79$_{2.003}$ & 65.85$_{2.872}$ & 64.31$_{4.129}$ &  86.09$_{3.304}$  & 76.02$_{5.041}$ & 33.69 & 20.87 & 6524\\ 
Att-UNet~\citep{schlemper2019attention} & VGG-16 & \underline{80.09}$_{1.372}$ & \underline{67.55}$_{1.750}$ & \underline{66.79}$_{2.091}$ & 79.68$_{2.063}$  & 66.67$_{2.823}$ & 7.88  & 43.35 & 2948\\
R50-\scriptsize{AttUNet}~\citep{schlemper2019attention} & ResNet-50 & 78.33$_{2.661}$ & 65.24$_{3.442}$ & 63.40$_{4.825}$ & 85.91$_{3.017}$  & 76.30$_{3.401}$ & 33.25 & 49.25 & 3546\\ 
R50-ViT~\citep{dosovitskiy2020image} & ResNet-50 & 75.25$_{1.229}$ & 60.97$_{1.486}$ & 56.91$_{2.316}$ & 77.87$_{1.036}$  & 64.39$_{1.427}$ & 110.62 & 26.91 & 8622\\
UNETR~\citep{hatamizadeh2022unetr} & ViT-B16 & 74.48$_{3.772}$ & 60.13$_{4.532}$ & 54.60$_{7.829}$ & 77.83$_{0.617}$  & 64.10$_{0.748}$ & 87.51 & 26.41 & 6004\\
Swin-UNETR~\citep{hatamizadeh2022swin} & Swin-T & 75.19$_{2.784}$ & 60.92$_{3.274}$ & 56.93$_{5.573}$ & 75.03$_{1.805}$  & 60.35$_{2.326}$ & 6.29  & 4.86 & 4616\\
\cdashline{1-10}[1.5pt/1.5pt]
MedSAM~\citep{ma2023segment} &   ViT-B16 &  73.84$_{3.188}$ & 59.13$_{3.814}$ & 54.78$_{6.200}$ & \underline{86.44}$_{1.428}$  & \underline{76.55}$_{2.121}$ & 93.74  & 28.56 & 6644\\
\cdashline{1-10}[1.5pt/1.5pt]
CTO (Ours) & CNNs-ViT & \textbf{80.68}$_{1.578}$ & \textbf{67.97}$_{2.113}$ & \textbf{67.17}$_{2.466}$ & \textbf{88.98}$_{1.901}$ & \textbf{80.55}$_{3.003}$ & 62.22 & 22.70 & 4104\\
\hline
\hline
\end{tabular}}
\vspace{-4mm}
\end{table*}
\begin{table*}[!t]
\small
\caption{Result comparisons (\%) with the state-of-the-art methods on BTCV~\citep{irshad2022improved}.}
\centering
\renewcommand\arraystretch{1.1}
\setlength{\tabcolsep}{5pt}{
\begin{tabular}{ r | c c | 
>{\centering\arraybackslash}p{1cm} 
>{\centering\arraybackslash}p{1cm} 
>{\centering\arraybackslash}p{1cm} 
>{\centering\arraybackslash}p{1cm} 
>{\centering\arraybackslash}p{1cm} 
>{\centering\arraybackslash}p{1cm} 
>{\centering\arraybackslash}p{1cm} 
>{\centering\arraybackslash}p{1cm}} 
\hline \hline
Methods & mDice $\uparrow$ & HD $\downarrow$ & Aorta & \scriptsize{Gallbladder} &\scriptsize{Kidney(L)} & \scriptsize{Kidney(R)} & Liver & Pancreas & Spleen & Stomach \\
\hline
V-Net~\citep{milletari2016v}     & 68.81        & -            & 75.34          & 51.87              & 77.10             & 80.75             & 87.84           & 40.05            & 80.56          & 56.98           \\
DARR~\citep{fu2020domain}      & 69.77        & -            & 74.74           & 53.77             & 72.31            & 73.24             & 94.08           & 54.18            & 89.90           & 45.96            \\
U-Net~\citep{ronneberger2015u}   & 76.85 & 39.70 & 89.07 & \textbf{69.72} & 77.77 & 68.60 & 93.43 & 53.98 & 86.67 & 75.58 \\
R50-UNet~\citep{ronneberger2015u}             & 74.68        & 36.87        & 84.18      & 62.84          & 79.19         & 71.29          & 93.35        & 48.23       & 84.41       & 73.92        \\
Att-UNet~\citep{schlemper2019attention} & 77.77 & 36.02 & \textbf{89.55} & \underline{68.88} & 77.98 & 71.11 & 93.57 & 58.04 & 87.30 & 75.75 \\
R50-\scriptsize{AttUNet}~\citep{schlemper2019attention} & 75.57   & 36.97        & 55.92      & 63.91          & 79.20         & 72.71          & 93.56       & 49.37        & 87.19       & 74.95        \\
R50-ViT~\citep{dosovitskiy2020image}    & 71.29        & 32.87        & 73.73      & 55.13          & 75.80         & 72.20         & 91.51       & 45.99        & 81.99      & 73.95        \\
TransUNet~\citep{chen2021transunet} & 77.48        & 31.69        & 87.23        & 63.13          & 81.87          & 77.02          & 94.08      & 55.86         & 85.08      & 75.62        \\ 
SwinUNet~\citep{cao2021swin} & 79.12 & 21.55 & 85.47 & 66.53 & 83.28 & 79.61 & 94.29 & 56.58 & 90.66 & 76.60 \\
UNETR++~\citep{shaker2024unetr} & \textbf{84.54} & \textbf{17.89} & \textbf{90.65} & 68.64 & \textbf{87.59} & \textbf{87.00} & \textbf{95.64} & \textbf{74.58} & \underline{90.70} &  \underline{81.48}  \\
\cdashline{1-11}[1.5pt/1.5pt]
MedSAM~\citep{ma2023segment} &    68.04      & -            &    53.62  &      61.87        & 72.86     &  67.96    &   77.55     &    \underline{69.36}     & 73.01           &   69.10          \\
\cdashline{1-11}[1.5pt/1.5pt]
CTO(Ours) & \underline{80.94} & \underline{19.45} & 86.72 & 64.68 & \underline{84.56} & \underline{80.82} & \underline{95.01} & 61.66 & \textbf{90.88} & \textbf{83.16} \\
\hline \hline
\end{tabular}}
\label{tab6}
\vspace{-4mm}
\end{table*}
\begin{table*}[!h]
    \small
    \caption{Result comparisons (\%) with the state-of-the-art methods on MSD BraTS~\citep{antonelli2022medical}.}
    \centering
    \renewcommand\arraystretch{1.1}
    \setlength{\tabcolsep}{12pt}{
    \begin{tabular}{r|ccccccccc}
    \hline \hline
    \multirow{2}{*}{Method}    & \multicolumn{2}{c}{Mean} & \multicolumn{2}{c}{WT} & \multicolumn{2}{c}{ET} & \multicolumn{2}{c}{TC} \\
    & HD~$\downarrow$  & Dice~$\uparrow$ & HD~$\downarrow$  & Dice~$\uparrow$  & HD~$\downarrow$ & Dice~$\uparrow$ & HD~$\downarrow$ & Dice~$\uparrow$       \\ 
    \hline 
    V-Net~\citep{milletari2016v} & 7.50  & 76.38   &  7.16  & 83.06  &  8.43  &  71.13 & 6.92 & 74.94 \\ 
    U-Net~\citep{ronneberger2015u} & 5.82  & 83.33   &  6.25  & 90.09  & 4.59   & 78.18  & 6.62 & 81.73  \\ 
    R50-UNet~\citep{ronneberger2015u} & 5.47  & 83.10   & 5.50   & 90.08  &  4.71  & 78.04  & 6.22 & 81.18 \\ 
    Att-UNet~\citep{schlemper2019attention} & 5.80  & 83.08 & 6.70 & 89.98 & 4.99 & 77.53 & \underline{5.74} & 81.75   \\ 
    R50-AttUNet~\citep{schlemper2019attention} & 6.81  &  83.11  &  5.80  & 90.47  &  7.58  & 76.42  & 7.06 & 82.44  \\ 
    R50-ViT~\citep{dosovitskiy2020image} & 5.95  &  82.82  & \underline{5.45}    &  89.72  & 5.50   &  77.13 & 6.91  & 81.59 \\
    TransUNet~\citep{chen2021transunet} & 5.43 & 83.63 & \textbf{4.93}  & \underline{90.55} & 4.96 & 77.90 & 6.30 & 82.42 \\ 
    SwinUNet~\citep{cao2021swin} & \textbf{5.12}  &  \underline{84.10}  &  5.61  &  90.38 &  \textbf{3.53}  & \textbf{79.87}  & 6.23 & \underline{82.57}  \\ 
    \cdashline{1-9}[1.5pt/1.5pt]
    MedSAM~\citep{ma2023segment} & 6.30  & 80.58   & 6.74   & 89.57  & 5.46   & 71.55  & 6.70 & 80.62 \\
    \cdashline{1-9}[1.5pt/1.5pt]
    CTO (ours) & \underline{5.42} & \textbf{84.62}  & 6.43  & \textbf{90.76} & \underline{4.56} & \underline{79.77}  & \textbf{5.27}  & \textbf{83.33} \\ 
    \hline \hline
    \end{tabular}}
\end{table*}

\myparagraph{Superiority of Boundary-Guided Decoder.}
We have carefully selected SOTA boundary-enhanced MedISeg methods to compare with our boundary-guided decoder. 
These methods include BATR~\citep{wang2021boundary}, XBound-Former~\citep{wang2023xbound}, PEE~\citep{wang2022boundary}, Boundary Loss~\citep{kervadec2019boundary}, and Boundary DoU Loss~\citep{sun2023boundary}. 
Specifically, we deployed different boundary detection methods on the same baseline model (\ie, CNNs + StitchViT), and the experimental results are shown in Table~\ref{tab2}. 
Notably, the incorporation of these boundary detection methods into the baseline model proves beneficial in further improving the overall performance of the model. 
This emphasizes the significance of incorporating boundary information in the context of MedISeg.

Furthermore, our CTO demonstrates the most significant performance improvement compared to the other methods, while maintaining a comparable increase in computational overheads. 
This not only highlights the effectiveness of the CTO, but also underscores its efficiency in terms of computational resources.

\myparagraph{Effectiveness of Each Component.} 
In this study, we conduct an extensive experimental analysis to compare the performance of various variants of CTO on the ISIC 2018 dataset, as documented in \cite{codella2019skin}. 
The evaluated variants encompass CNNs, StitchViT, CBM, BEM, and BIM.
CBM indicates the convolutional boundary module, which is specifically designed to extract boundary features without employing the Sobel operator. 
Our results, presented in Table~\ref{tab3}, unequivocally demonstrate the substantial performance enhancements that each of these components can offer over the baseline model. 
Notably, when incorporated into the CNNs-based baseline, the +StitchViT, +CBM, +BEM, and +BIM variants yield Dice score improvements of 0.42\%, 0.56\%, 1.01\%, and 1.45\%, respectively. 
Additionally, these components exhibit improvements in IoU performance, with gains of 0.46\%, 0.57\%, 1.10\%, and 1.57\% on the CNNs-based baseline, respectively. These compelling findings underscore the effectiveness of the evaluated CTO variants in enhancing the segmentation performance of the baseline model.

Furthermore, our results reveal that the addition of a CNNs model atop StitchViT yields superior performance compared to ViT alone. 
This observation suggests that the fusion of different models can yield enhanced performance compared to relying on a single model.
Of particular importance, our findings highlight the remarkable improvement achieved by the BIM variant, with a notable increase of 1.45\% in Dice score and 1.51\% in IoU score. 
This observation emphasizes the pivotal role played by BIM in enhancing segmentation performance in the realm of medical imaging applications. 
The results presented in this ablation study unequivocally demonstrate the substantial improvements in baseline performance achieved by the evaluated CTO variants on the ISIC 2018 dataset. These findings underscore the significance of these components in improving performance in medical imaging applications and suggest that well-designed combinations of different models can lead to further advancements in performance.

\begin{figure*}[!t]
\centering
\includegraphics[width=.95\textwidth]{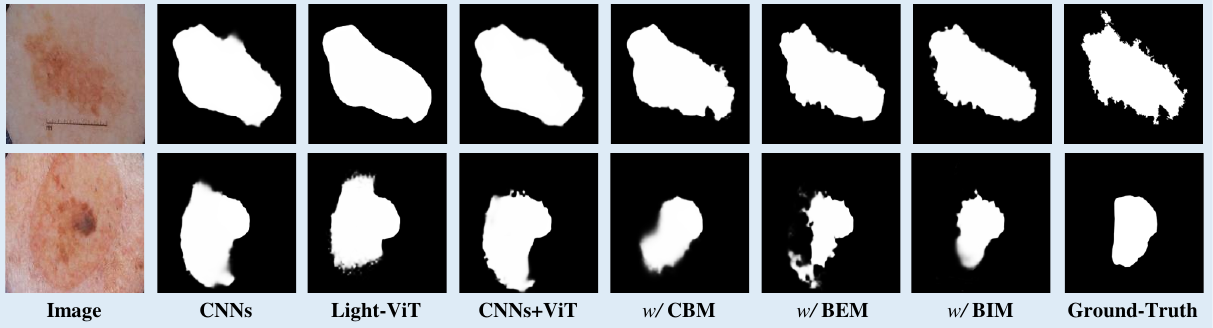}
\caption{Visualized results of the ablation study on ISIC 2018~\cite{codella2019skin}. ``\emph{w/}'' denotes with the corresponding module implementation.}
\label{fig:3}
\end{figure*}
\begin{figure*}[t]
\centering
\includegraphics[width=.95\textwidth]{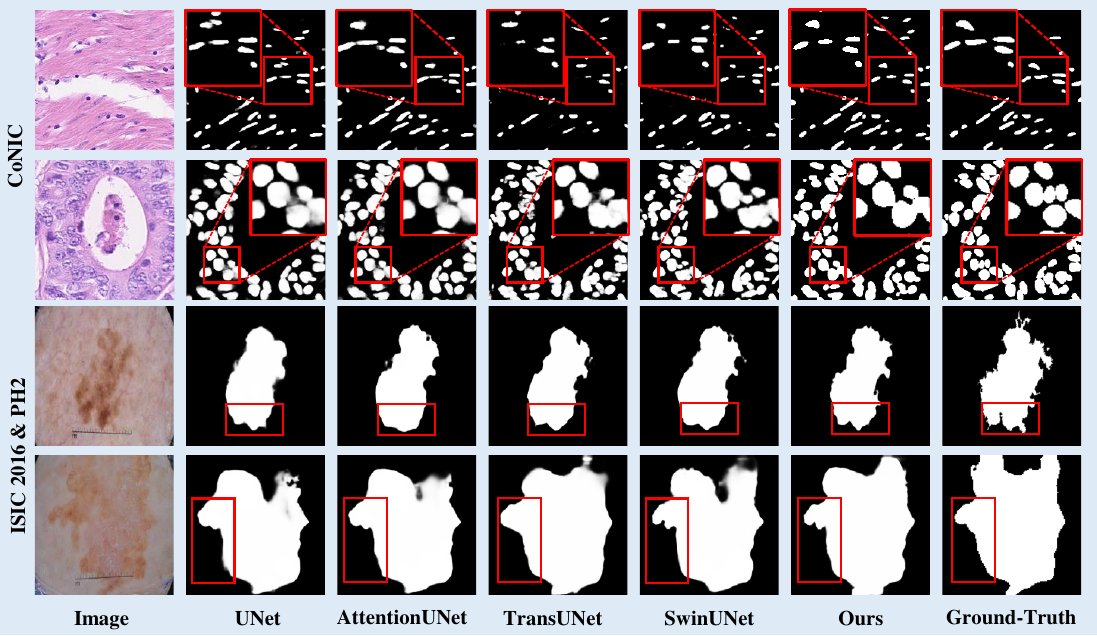}
\caption{{Visualizations on CoNIC~\cite{graham2021conic} and ISIC 2016~\cite{gutman2016skin}\&PH2~\cite{mendoncca2013ph}. The \textcolor{red}{red} frames highlight the improved regions.}}
\label{fig:5}
\vspace{-4mm}
\end{figure*}
\begin{figure*}[t]
\centering
\includegraphics[width=.99\textwidth]{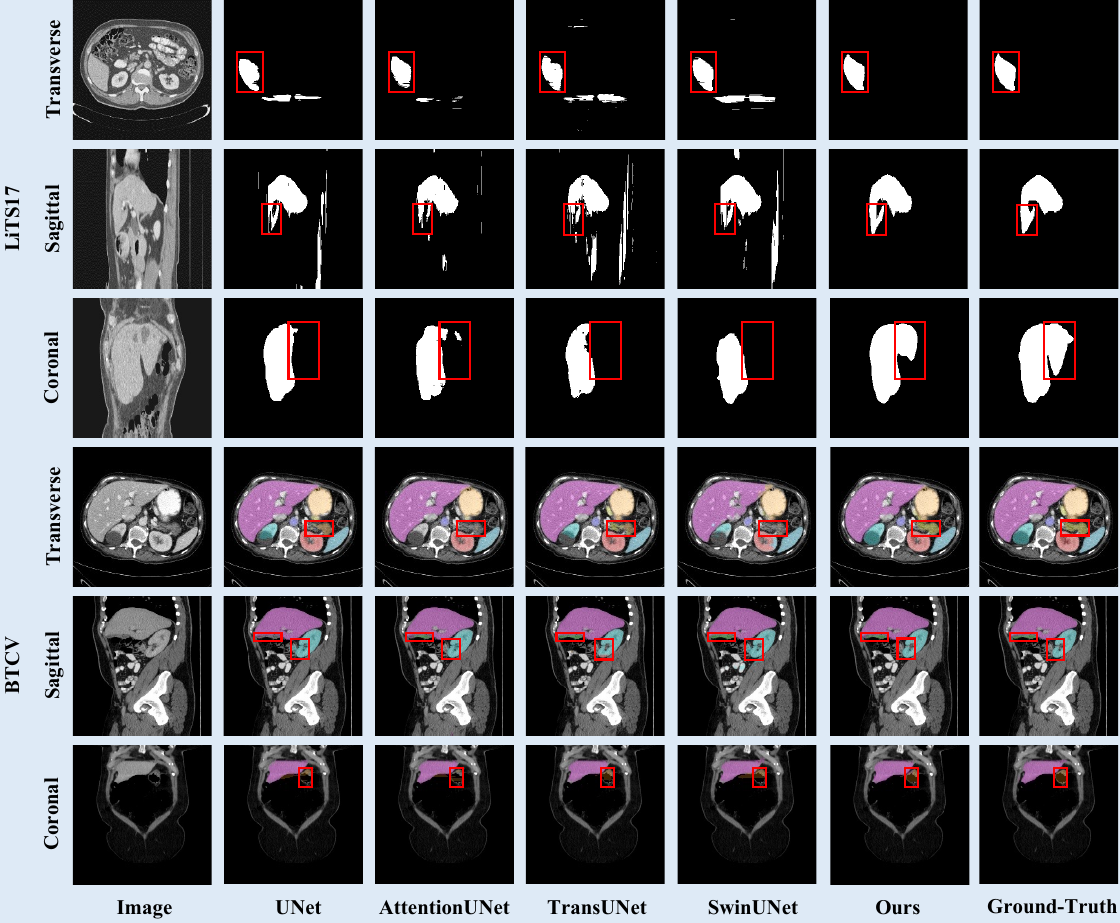}
\caption{Visualizations on LiTS17~\cite{bilic2019liver}  and BTCV~\cite{vaswani2017attention} in transverse, sagittal, and coronal plane. The \textcolor{red}{red} frames highlight the improved regions.}
\label{fig:lits_btcv}
\vspace{\vs}
\end{figure*}
\begin{figure*}[t]
\centering
\includegraphics[width=.99\textwidth]{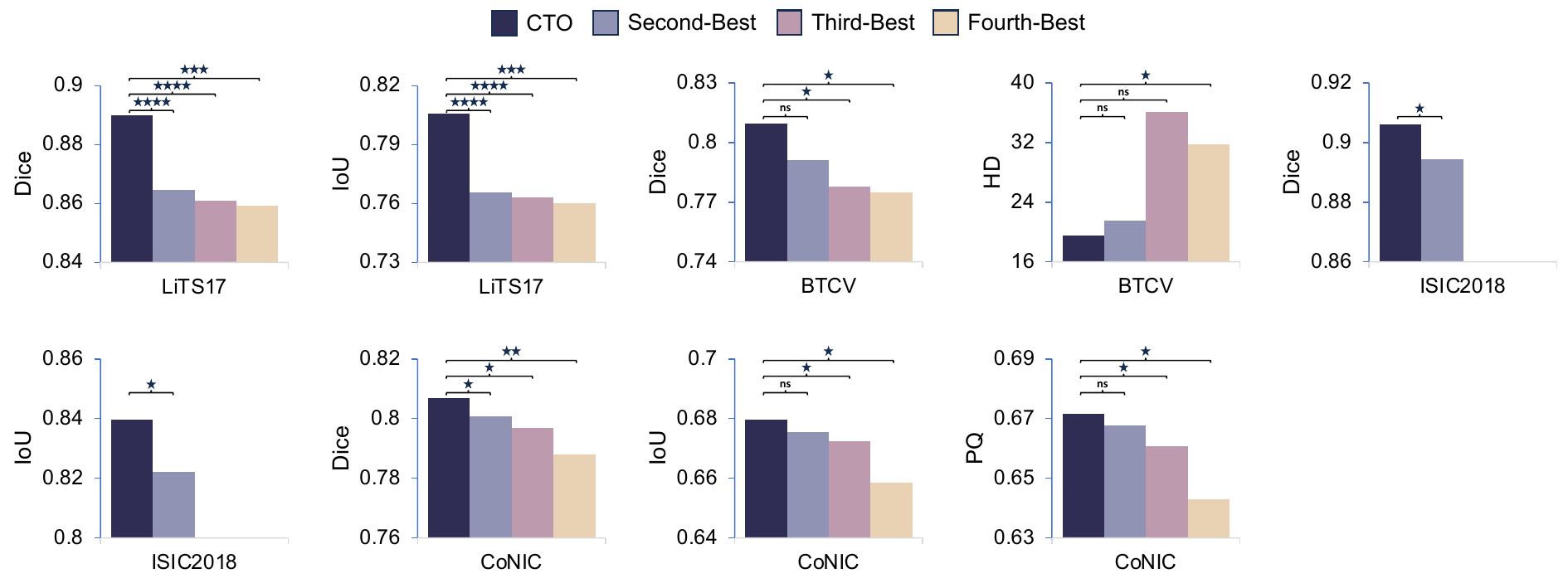}
\caption{{Statistical analyses in terms of each quantitative metric in the manuscript between CTO and the second-best, third-best and fourth-best performing methods in ISIC2018, BTCV, CoNIC and LiTS17. $p$-values are calculated using paired t-test with the Bonferroni correction and referred as follows: ``\textit{ns}'' on the plot refers to the $p$-value in the range: $0.05 < p <= 1$; $^{*}$ refers to the $p$-value in the range: $0.01 < p <= 0.05$; $^{**}$ refers to the $p$-value in the range: $0.001 < p <= 0.01$; $^{***}$ refers to the $p$-value in the range: $0.0001 < p <= 0.001$; $^{****}$ refers to the $p$-value in the range: $p <= 0.0001$.}}
\label{fig:statistics}
\vspace{\vs}
\end{figure*}

\subsection{Comparisons with State-of-the-Arts}
{To demonstrate the exceptional performance and versatility of our proposed CTO, we conduct comparisons with SOTA methods on various MedISeg benchmarks encompassing different modalities, including dermatologic images (\ie, ISIC 2016~\citep{gutman2016skin}, PH2~\citep{mendoncca2013ph}, and ISIC 2018~\citep{codella2019skin}), histopathological images (CoNIC~\citep{graham2021conic}), and radiological images (LiTS17~\citep{bilic2019liver}, MSD BraTS~\citep{antonelli2022medical}, and BTCV~\citep{vaswani2017attention}).}

On the ISIC 2016  and PH2 datasets, we compare our CTO with five related methods, namely, SSLS~\citep{ahn2015automated}, MSCA~\citep{bi2016automated}, FCN~\citep{long2015fully}, MFCN~\citep{bi2017dermoscopic}, and SBPSeg~\citep{lee2020structure}. 
The results, presented in Table~\ref{tab4}, exhibit notable achievements by our CTO, surpassing the previous SOTA SBPSeg~\citep{lee2020structure}  by 0.72\% and 2.30\% in Dice and IoU scores, respectively, with an outstanding performance of 92.56\% in Dice and 86.60\% in IoU.
On the ISIC 2018 dataset, our CTO achieves remarkable results with 90.6\% in Dice and 84.0\% in IoU through 5-fold cross-validation. These scores outperform the SOTA TransUNet~\citep{chen2021transunet} by 1.2\% and 1.8\%, respectively, further demonstrating the robust learning capacity and generalizability of our proposed CTO. 
These results also underscore the pivotal role of boundary information in lesion segmentation.

{
Moving on to the CoNIC dataset, our CTO consistently outperforms other methods through five-fold cross validation, as evidenced by the results presented on the left side of Table~\ref{tab5}. Notably, our method excels over the current SOTA Att-UNet~\citep{schlemper2019attention} by 0.59\% in Dice, 0.42\% in IoU, and 0.38\% in PQ, respectively. 
}These compelling results on ISIC 2016, PH2, ISIC 2018, and CoNIC validate the superiority of our method in the realm of 2D medical image segmentation tasks.

{In addition to the 2D MedISeg, we have also conducted experiments on 3D MedISeg, utilizing the LiTS17, MSD BraTS, BTCV datasets. 
On LiTS17, as depicted on the right side of Table~\ref{tab5}, our method achieves an average of 88.98\% in Dice and 80.55\% in IoU across five folds, surpassing the second-best performing method, R50-UNet~\citep{ronneberger2015u} by 2.89\% and 4.53\%, respectively.
On MSD BraTS, our CTO achieves an average Dice score of 84.62\%, outperforming the second-best performing method, SwinUNet\citep{cao2021swin}, by 0.52\%. In terms of HD, our method achieves an average of 5.42 mm, positioning it as the second-best performer among the compared methods.}

{Turning our attention to BTCV, as presented in Table~\ref{tab6}, our proposed CTO achieves remarkable Dice scores of 80.94\%, ranking the second-best performance among the methods.
Particularly noteworthy is the significant improvement observed for organs with blurry boundaries, such as ``spleen" and ``stomach," where our model achieves gains of 0.18\% and 1.68\% in Dice, respectively. 
Our method is inferior to the UNETR++ in the task of 3D organ segmentation, we believe that the performance of our method can be further improved by extending the model to a 3D version.
Moreover, our CTO exhibits competitive performance improvements while upholding a comparable computational load.
The final model configuration of CTO comprises 62.22M parameters, 22.70G FLOPs, and 4104 MB of GPU memory.}

{\subsection{Statistical Analysis}
To compare the disparities between our methodology and the current leading methods, we conduct a statistical significance analysis using the paired t-test with the Bonferroni correction. The results are visually represented with bars atop the box plots, where significance levels are denoted as follows: non-significant (\textit{ns}, $p>0.05$), significant: * ($0.01<p<0.05$), ** ($0.001<p<0.01$), *** ($p<0.001$). The statistical analysis outcomes are illustrated in Fig.~\ref{fig:statistics}.}

{Across the ISIC 2018, CoNIC, and LiTS datasets, our approach demonstrates a significant superiority over the second-best methods (TransUNet, Attention UNet, and MedSAM). Notably, on the BTCV dataset, our method exhibits a remarkable enhancement in the Dice score, increasing from 77.77\% to 80.94\% when compared to the third-best performing method (Attention UNet). Furthermore, in terms of HD, CTO showcases a substantial decrease from 36.02 to 19.45 when juxtaposed with the fourth-best performing method (Attention UNet). These findings affirm that CTO consistently enhances performance significantly across all datasets compared to the majority of state-of-the-art methods.}

\subsection{Comparisons with Large Vision Models}
Recently, the emergence of large vision models trained on extensive medical image datasets has garnered significant attention and is being hailed as the future direction for MedISeg. 
In light of this, in this section, we compare our method with the large vision model in medical images, namely, Segment Anything in Medical Images (MedSAM)~\citep{ma2023segment}. 

Examining the rows labeled ``MedSAM" in Table~\ref{tab4} to Table~\ref{tab6}, our method consistently outperforms MedSAM on specific datasets. 
Moreover, our CTO exhibits the potential for fewer parameters and more lightweight computations in comparison to MedSAM. 
This observation suggests that while MedSAM may possess robust learning capabilities, its performance on specific datasets remains unsatisfactory. 
Therefore, it is worth considering the development of more efficient and effective dataset-specific models by combining the downstream application of large MedSAM models on specific datasets.

\section{Qualitative Results}
\subsection{Comparisons with the Baseline Model.}
The primary objective of this study is to investigate the performance of various models on the ISIC 2018 dataset~\citep{codella2019skin}. 
Specifically, we aim to evaluate the efficacy of CNNs, StitchViT, and their combination, as well as the impact of incorporating CBM, BEM, and BIM into the combined model.
Our results, as depicted in Figure~\ref{fig:3}, unequivocally demonstrate that CNN-based models excel at capturing local information pertaining to foreground objects. 
Conversely, ViT-based models excel at retrieving global information related to foreground objects. 
However, both of these models exhibit the drawback of incomplete and imprecise object boundaries.
On the other hand, the CNNs+ViT approach not only effectively captures local object information to enhance the smoothness of local masks but also ensures the overall completeness of objects.
Furthermore, our results highlight that the addition of CBM, BEM, and BIM to the CNNs+ViT model yields a significant improvement in both boundary delineation and object completeness. 
This finding underscores the crucial role played by the explicit boundary features learned by our model in the segmentation of foreground objects.
Overall, the experimental results demonstrate the effectiveness of the evaluated models and emphasize the significance of integrating explicit boundary features to enhance segmentation in medical imaging applications.

\subsection{Comparisons with State-of-the-Arts.}
{Figure~\ref{fig:5} provides qualitative comparisons of the results on the CoNIC~\citep{graham2021conic} and  ISIC 2016~\citep{gutman2016skin} \& PH2~\citep{mendoncca2013ph}, showcasing the effectiveness of our proposed CTO method.
Our results demonstrate the remarkable ability of our CTO method to accurately delineate object contours, particularly for nuclei with diverse shapes and sizes, and even for nuclei objects with blurred boundaries. 
This observation underscores the significance of incorporating boundary information to enhance segmentation performance in medical image analysis tasks.}

{Moreover, Figure~\ref{fig:lits_btcv} provides qualitative comparisons of the results on the LiTS17~\cite{bilic2019liver} and BTCV~\cite{vaswani2017attention} in coronal, sagittal
and transverse plane. Our results establish the superiority of our CTO method over other existing methods. For instance, in all three views of the LiTS17 dataset, our method successfully segments more comprehensive boundaries, leading to a more accurate representation of liver. Moreover, in the coronal planes of images in the BTCV dataset, CTO accurately distinguishes boundaries between the
pancreas and left kidney, while other methods exhibit under-segmentation of pancreas. Similarly, in the sagittal and transverse planes of images in the BTCV dataset, CTO effectively segments more accurate boundaries of right kidney and stomach compared to other methods.} 

{These findings provide compelling evidence for the effectiveness of our proposed method in addressing medical image segmentation challenges. The experimental results not only highlight the importance of boundary information in medical image segmentation tasks but also underscore the superiority of our CTO method in comparison to other existing methods.}
\section{Conclusion}
In this paper, we proposed a new network architecture, named CTO, specifically tailored for medical image segmentation.
The proposed architecture not only achieves a superior balance between recognition accuracy and computational efficiency but also outperforms existing advanced segmentation architectures.
The primary contribution of this paper lies in the ingenious use of intermediate feature maps to synthesize a high-quality boundary supervision mask, eliminating the need for additional external information. 
To this end, we designed the CTO network, which seamlessly integrates three crucial components -- convolution, Transformer, and edge detection operator -- into a unified framework.
Through extensive experimentation on seven publicly available datasets, we provide compelling evidence of the unmatched superiority of CTO over state-of-the-art methods. 
Furthermore, we demonstrate the exceptional effectiveness of each individual component within our architecture.
The limitations of the proposed method are as follows:
1) This study does not explore the use of 3D convolutions, which could better capture spatial relationships in volumetric data, potentially improving segmentation accuracy.
2) We did not investigate the impact of domain shift between different medical imaging centers, which can affect model generalization.
3) The proposed method is not yet fully optimized for real-time applications, which may limit its practical use in clinical settings.
4) The model does not leverage pre-trained knowledge from large models, which could enhance feature learning and overall performance.
{The future work will concentrate on extending the concept of a dual-stream encoder to encompass diverse advanced backbone architectures, thereby elevating the effectiveness of the proposed approach.
Additionally, we aim to explore the potential adaptation of CTO to a three-dimensional framework, which holds immense promise for revolutionizing medical image analysis tasks in real-world clinical practice.}
\section*{Declaration of generative AI and AI-assisted technologies in the writing process}
During the preparation of this work the authors used ChatGPT 3.5 Turbo in order to improve language and readability. After using this tool, the authors reviewed and edited the content as needed and take full responsibility for the content of the publication.
\bibliographystyle{model2-names.bst}
\biboptions{authoryear}
\bibliography{refs}

\begin{thebibliography}{78}
\expandafter\ifx\csname natexlab\endcsname\relax\def\natexlab#1{#1}\fi
\providecommand{\url}[1]{\texttt{#1}}
\providecommand{\href}[2]{#2}
\providecommand{\path}[1]{#1}
\providecommand{\DOIprefix}{doi:}
\providecommand{\ArXivprefix}{arXiv:}
\providecommand{\URLprefix}{URL: }
\providecommand{\Pubmedprefix}{pmid:}
\providecommand{\doi}[1]{\href{http://dx.doi.org/#1}{\path{#1}}}
\providecommand{\Pubmed}[1]{\href{pmid:#1}{\path{#1}}}
\providecommand{\bibinfo}[2]{#2}
\ifx\xfnm\relax \def\xfnm[#1]{\unskip,\space#1}\fi
\bibitem[{Ahn et~al.(2015)Ahn, Bi, Jung, Kim, Li, Fulham and Feng}]{ahn2015automated}
\bibinfo{author}{Ahn, E.}, \bibinfo{author}{Bi, L.}, \bibinfo{author}{Jung, Y.H.}, \bibinfo{author}{Kim, J.}, \bibinfo{author}{Li, C.}, \bibinfo{author}{Fulham, M.}, \bibinfo{author}{Feng, D.D.}, \bibinfo{year}{2015}.
\newblock \bibinfo{title}{Automated saliency-based lesion segmentation in dermoscopic images}, in: \bibinfo{booktitle}{IEEE Engineering in Medicine and Biology Society}, pp. \bibinfo{pages}{3009--3012}.
\newblock \DOIprefix\doi{10.1109/EMBC.2015.7319025}.
\bibitem[{Antonelli et~al.(2022)Antonelli, Reinke, Bakas, Farahani, Kopp-Schneider, Landman, Litjens, Menze, Ronneberger, Summers et~al.}]{antonelli2022medical}
\bibinfo{author}{Antonelli, M.}, \bibinfo{author}{Reinke, A.}, \bibinfo{author}{Bakas, S.}, \bibinfo{author}{Farahani, K.}, \bibinfo{author}{Kopp-Schneider, A.}, \bibinfo{author}{Landman, B.A.}, \bibinfo{author}{Litjens, G.}, \bibinfo{author}{Menze, B.}, \bibinfo{author}{Ronneberger, O.}, \bibinfo{author}{Summers, R.M.}, et~al., \bibinfo{year}{2022}.
\newblock \bibinfo{title}{The medical segmentation decathlon}.
\newblock \bibinfo{journal}{Nature Communications} \bibinfo{volume}{13}, \bibinfo{pages}{4128}.
\newblock \DOIprefix\doi{10.1038/s41467-022-30695-9}.
\bibitem[{Bi et~al.(2016)Bi, Kim, Ahn, Feng and Fulham}]{bi2016automated}
\bibinfo{author}{Bi, L.}, \bibinfo{author}{Kim, J.}, \bibinfo{author}{Ahn, E.}, \bibinfo{author}{Feng, D.}, \bibinfo{author}{Fulham, M.}, \bibinfo{year}{2016}.
\newblock \bibinfo{title}{Automated skin lesion segmentation via image-wise supervised learning and multi-scale superpixel based cellular automata}, in: \bibinfo{booktitle}{IEEE International Symposium on Biomedical Imaging}, pp. \bibinfo{pages}{1059--1062}.
\newblock \DOIprefix\doi{10.1109/ISBI.2016.7493448}.
\bibitem[{Bi et~al.(2017)Bi, Kim, Ahn, Kumar, Fulham and Feng}]{bi2017dermoscopic}
\bibinfo{author}{Bi, L.}, \bibinfo{author}{Kim, J.}, \bibinfo{author}{Ahn, E.}, \bibinfo{author}{Kumar, A.}, \bibinfo{author}{Fulham, M.}, \bibinfo{author}{Feng, D.}, \bibinfo{year}{2017}.
\newblock \bibinfo{title}{Dermoscopic image segmentation via multistage fully convolutional networks}.
\newblock \bibinfo{journal}{IEEE Transactions on Biomedical Engineering} \bibinfo{volume}{64}, \bibinfo{pages}{2065--2074}.
\newblock \DOIprefix\doi{10.1109/TBME.2017.2712771}.
\bibitem[{Bilic et~al.(2023)Bilic, Christ, Li, Vorontsov, Ben-Cohen, Kaissis, Szeskin, Jacobs, Mamani, Chartrand et~al.}]{bilic2019liver}
\bibinfo{author}{Bilic, P.}, \bibinfo{author}{Christ, P.}, \bibinfo{author}{Li, H.B.}, \bibinfo{author}{Vorontsov, E.}, \bibinfo{author}{Ben-Cohen, A.}, \bibinfo{author}{Kaissis, G.}, \bibinfo{author}{Szeskin, A.}, \bibinfo{author}{Jacobs, C.}, \bibinfo{author}{Mamani, G.E.H.}, \bibinfo{author}{Chartrand, G.}, et~al., \bibinfo{year}{2023}.
\newblock \bibinfo{title}{The liver tumor segmentation benchmark ({LiTS})}.
\newblock \bibinfo{journal}{Medical Image Analysis} \bibinfo{volume}{84}, \bibinfo{pages}{102680}.
\newblock \DOIprefix\doi{10.1016/j.media.2022.102680}.
\bibitem[{Bokhovkin and Burnaev(2019)}]{bokhovkin2019boundary}
\bibinfo{author}{Bokhovkin, A.}, \bibinfo{author}{Burnaev, E.}, \bibinfo{year}{2019}.
\newblock \bibinfo{title}{Boundary loss for remote sensing imagery semantic segmentation}, in: \bibinfo{booktitle}{International Symposium on Neural Networks}, \bibinfo{publisher}{Lecture Notes in Computer Science, 11555, Springer}. pp. \bibinfo{pages}{388--401}.
\newblock \DOIprefix\doi{10.1007/978-3-030-22808-8_38}.
\bibitem[{Burger and Burge(2022)}]{burger2022digital}
\bibinfo{author}{Burger, W.}, \bibinfo{author}{Burge, M.J.}, \bibinfo{year}{2022}.
\newblock \bibinfo{title}{Digital Image Processing: An Algorithmic Introduction}.
\newblock \bibinfo{publisher}{Springer Nature}.
\newblock \DOIprefix\doi{10.1007/978-3-031-05744-1}.
\bibitem[{Cao et~al.(2021)Cao, Wang, Chen, Jiang, Zhang, Tian and Wang}]{cao2021swin}
\bibinfo{author}{Cao, H.}, \bibinfo{author}{Wang, Y.}, \bibinfo{author}{Chen, J.}, \bibinfo{author}{Jiang, D.}, \bibinfo{author}{Zhang, X.}, \bibinfo{author}{Tian, Q.}, \bibinfo{author}{Wang, M.}, \bibinfo{year}{2021}.
\newblock \bibinfo{title}{Swin-{UNet}: Unet-like pure {Transformer} for medical image segmentation}, in: \bibinfo{booktitle}{European Conference on Computer Vision Workshop}, \bibinfo{publisher}{Lecture Notes in Computer Science, vol 13803. Springer, Cham}. pp. \bibinfo{pages}{205--218}.
\newblock \DOIprefix\doi{10.1007/978-3-031-25066-8_9}.
\bibitem[{Chen et~al.(2018)Chen, Dou, Yu, Qin and Heng}]{chen2018voxresnet}
\bibinfo{author}{Chen, H.}, \bibinfo{author}{Dou, Q.}, \bibinfo{author}{Yu, L.}, \bibinfo{author}{Qin, J.}, \bibinfo{author}{Heng, P.A.}, \bibinfo{year}{2018}.
\newblock \bibinfo{title}{{VoxResNet}: Deep voxelwise residual networks for brain segmentation from {3D} {MR} images}.
\newblock \bibinfo{journal}{NeuroImage} \bibinfo{volume}{170}, \bibinfo{pages}{446--455}.
\newblock \DOIprefix\doi{10.1016/j.neuroimage.2017.04.041}.
\bibitem[{Chen et~al.(2016)Chen, Qi, Yu and Heng}]{chen2016dcan}
\bibinfo{author}{Chen, H.}, \bibinfo{author}{Qi, X.}, \bibinfo{author}{Yu, L.}, \bibinfo{author}{Heng, P.A.}, \bibinfo{year}{2016}.
\newblock \bibinfo{title}{{DCAN}: deep contour-aware networks for accurate gland segmentation}, in: \bibinfo{booktitle}{Proceedings of the IEEE/CVF Conference on Computer Vision and Pattern Recognition}, pp. \bibinfo{pages}{1487--1496}.
\newblock \DOIprefix\doi{10.1109/CVPR.2016.273}.
\bibitem[{Chen et~al.(2024)Chen, Mei, Li, Lu, Yu, Wei, Luo, Xie, Adeli, Wang et~al.}]{chen2021transunet}
\bibinfo{author}{Chen, J.}, \bibinfo{author}{Mei, J.}, \bibinfo{author}{Li, X.}, \bibinfo{author}{Lu, Y.}, \bibinfo{author}{Yu, Q.}, \bibinfo{author}{Wei, Q.}, \bibinfo{author}{Luo, X.}, \bibinfo{author}{Xie, Y.}, \bibinfo{author}{Adeli, E.}, \bibinfo{author}{Wang, Y.}, et~al., \bibinfo{year}{2024}.
\newblock \bibinfo{title}{{TransUNet}: Rethinking the u-net architecture design for medical image segmentation through the lens of {Transformers}}.
\newblock \bibinfo{journal}{Medical Image Analysis} \bibinfo{volume}{97}, \bibinfo{pages}{103280}.
\newblock \DOIprefix\doi{10.1016/j.media.2024.103280}.
\bibitem[{Chen et~al.(2017)Chen, Papandreou, Schroff and Adam}]{chen2017rethinking}
\bibinfo{author}{Chen, L.C.}, \bibinfo{author}{Papandreou, G.}, \bibinfo{author}{Schroff, F.}, \bibinfo{author}{Adam, H.}, \bibinfo{year}{2017}.
\newblock \bibinfo{title}{Rethinking atrous convolution for semantic image segmentation}.
\newblock \bibinfo{journal}{arXiv:1706.05587} .
\bibitem[{Chen et~al.(2021)Chen, Dong, Ji, Cao and Li}]{chen2021image}
\bibinfo{author}{Chen, X.}, \bibinfo{author}{Dong, C.}, \bibinfo{author}{Ji, J.}, \bibinfo{author}{Cao, J.}, \bibinfo{author}{Li, X.}, \bibinfo{year}{2021}.
\newblock \bibinfo{title}{Image manipulation detection by multi-view multi-scale supervision}, in: \bibinfo{booktitle}{Proceedings of the IEEE International Conference on Computer Vision}, pp. \bibinfo{pages}{14185--14193}.
\newblock \DOIprefix\doi{10.1109/ICCV48922.2021.01392}.
\bibitem[{Chu et~al.(2021)Chu, Tian, Wang, Zhang, Ren, Wei, Xia and Shen}]{chu2021twins}
\bibinfo{author}{Chu, X.}, \bibinfo{author}{Tian, Z.}, \bibinfo{author}{Wang, Y.}, \bibinfo{author}{Zhang, B.}, \bibinfo{author}{Ren, H.}, \bibinfo{author}{Wei, X.}, \bibinfo{author}{Xia, H.}, \bibinfo{author}{Shen, C.}, \bibinfo{year}{2021}.
\newblock \bibinfo{title}{Twins: Revisiting the design of spatial attention in vision {Transformer}s}.
\newblock \bibinfo{journal}{Advances in Neural Information Processing Systems} \bibinfo{volume}{34}, \bibinfo{pages}{9355--9366}.
\newblock \URLprefix \url{https://openreview.net/forum?id=5kTlVBkzSRx}.
\bibitem[{Codella et~al.(2019)Codella, Rotemberg, Tschandl, Celebi, Dusza, Gutman, Helba, Kalloo, Liopyris, Marchetti et~al.}]{codella2019skin}
\bibinfo{author}{Codella, N.}, \bibinfo{author}{Rotemberg, V.}, \bibinfo{author}{Tschandl, P.}, \bibinfo{author}{Celebi, M.E.}, \bibinfo{author}{Dusza, S.}, \bibinfo{author}{Gutman, D.}, \bibinfo{author}{Helba, B.}, \bibinfo{author}{Kalloo, A.}, \bibinfo{author}{Liopyris, K.}, \bibinfo{author}{Marchetti, M.}, et~al., \bibinfo{year}{2019}.
\newblock \bibinfo{title}{Skin lesion analysis toward melanoma detection 2018: A challenge hosted by the international skin imaging collaboration ({ISIC})}.
\newblock \bibinfo{journal}{arXiv:1902.03368} .
\bibitem[{Deng et~al.(2009)Deng, Dong, Socher, Li, Li and Fei-Fei}]{deng2009imagenet}
\bibinfo{author}{Deng, J.}, \bibinfo{author}{Dong, W.}, \bibinfo{author}{Socher, R.}, \bibinfo{author}{Li, L.J.}, \bibinfo{author}{Li, K.}, \bibinfo{author}{Fei-Fei, L.}, \bibinfo{year}{2009}.
\newblock \bibinfo{title}{{ImageNet}: A large-scale hierarchical image database}, in: \bibinfo{booktitle}{Proceedings of the IEEE/CVF Conference on Computer Vision and Pattern Recognition}, pp. \bibinfo{pages}{248--255}.
\newblock \DOIprefix\doi{10.1109/CVPR.2009.5206848}.
\bibitem[{Dosovitskiy et~al.(2020)Dosovitskiy, Beyer, Kolesnikov, Weissenborn, Zhai, Unterthiner, Dehghani, Minderer, Heigold, Gelly et~al.}]{dosovitskiy2020image}
\bibinfo{author}{Dosovitskiy, A.}, \bibinfo{author}{Beyer, L.}, \bibinfo{author}{Kolesnikov, A.}, \bibinfo{author}{Weissenborn, D.}, \bibinfo{author}{Zhai, X.}, \bibinfo{author}{Unterthiner, T.}, \bibinfo{author}{Dehghani, M.}, \bibinfo{author}{Minderer, M.}, \bibinfo{author}{Heigold, G.}, \bibinfo{author}{Gelly, S.}, et~al., \bibinfo{year}{2020}.
\newblock \bibinfo{title}{An image is worth 16x16 words: {Transformers} for image recognition at scale}, in: \bibinfo{booktitle}{International Conference on Learning Representations}, pp. \bibinfo{pages}{1909--1920}.
\newblock \URLprefix \url{https://openreview.net/forum?id=YicbFdNTTy}.
\bibitem[{Fan et~al.(2020)Fan, Ji, Sun, Cheng, Shen and Shao}]{fan2020camouflaged}
\bibinfo{author}{Fan, D.P.}, \bibinfo{author}{Ji, G.P.}, \bibinfo{author}{Sun, G.}, \bibinfo{author}{Cheng, M.M.}, \bibinfo{author}{Shen, J.}, \bibinfo{author}{Shao, L.}, \bibinfo{year}{2020}.
\newblock \bibinfo{title}{Camouflaged object detection}, in: \bibinfo{booktitle}{Proceedings of the IEEE/CVF Conference on Computer Vision and Pattern Recognition}, pp. \bibinfo{pages}{2777--2787}.
\newblock \DOIprefix\doi{10.1109/CVPR42600.2020.00285}.
\bibitem[{Fu et~al.(2020)Fu, Lu, Wang, Zhou, Shen, Fishman and Yuille}]{fu2020domain}
\bibinfo{author}{Fu, S.}, \bibinfo{author}{Lu, Y.}, \bibinfo{author}{Wang, Y.}, \bibinfo{author}{Zhou, Y.}, \bibinfo{author}{Shen, W.}, \bibinfo{author}{Fishman, E.}, \bibinfo{author}{Yuille, A.}, \bibinfo{year}{2020}.
\newblock \bibinfo{title}{Domain adaptive relational reasoning for {3D} multi-organ segmentation}, in: \bibinfo{booktitle}{International Conference on Medical Image Computing and Computer Assisted Intervention}, \bibinfo{publisher}{Lecture Notes in Computer Science, vol 12261. Springer, Cham}. pp. \bibinfo{pages}{656--666}.
\newblock \DOIprefix\doi{10.1007/978-3-030-59710-8_64}.
\bibitem[{Gao et~al.(2019)Gao, Cheng, Zhao, Zhang, Yang and Torr}]{gao2019res2net}
\bibinfo{author}{Gao, S.H.}, \bibinfo{author}{Cheng, M.M.}, \bibinfo{author}{Zhao, K.}, \bibinfo{author}{Zhang, X.Y.}, \bibinfo{author}{Yang, M.H.}, \bibinfo{author}{Torr, P.}, \bibinfo{year}{2019}.
\newblock \bibinfo{title}{{Res2Net}: A new multi-scale backbone architecture}.
\newblock \bibinfo{journal}{IEEE Transactions on Pattern Analysis and Machine Intelligence} \bibinfo{volume}{43}, \bibinfo{pages}{652--662}.
\newblock \DOIprefix\doi{10.1109/TPAMI.2019.2938758}.
\bibitem[{Gao et~al.(2021)Gao, Jin, Zhao, Dou and Heng}]{gao2021future}
\bibinfo{author}{Gao, X.}, \bibinfo{author}{Jin, Y.}, \bibinfo{author}{Zhao, Z.}, \bibinfo{author}{Dou, Q.}, \bibinfo{author}{Heng, P.A.}, \bibinfo{year}{2021}.
\newblock \bibinfo{title}{Future frame prediction for robot-assisted surgery}, in: \bibinfo{booktitle}{International Conference on Information Processing in Medical Imaging}, \bibinfo{publisher}{Lecture Notes in Computer Science, vol 12729. Springer, Cham}. pp. \bibinfo{pages}{533--544}.
\newblock \DOIprefix\doi{10.1007/978-3-030-78191-0_41}.
\bibitem[{Gao et~al.(2022)Gao, Zhou, Liu, Yan, Zhang and Metaxas}]{gao2022data}
\bibinfo{author}{Gao, Y.}, \bibinfo{author}{Zhou, M.}, \bibinfo{author}{Liu, D.}, \bibinfo{author}{Yan, Z.}, \bibinfo{author}{Zhang, S.}, \bibinfo{author}{Metaxas, D.N.}, \bibinfo{year}{2022}.
\newblock \bibinfo{title}{A data-scalable {Transformer} for medical image segmentation: architecture, model efficiency, and benchmark}.
\newblock \bibinfo{journal}{arXiv preprint arXiv:2203.00131} .
\bibitem[{Graham et~al.(2021a)Graham, El-Nouby, Touvron, Stock, Joulin, J{\'e}gou and Douze}]{graham2021levit}
\bibinfo{author}{Graham, B.}, \bibinfo{author}{El-Nouby, A.}, \bibinfo{author}{Touvron, H.}, \bibinfo{author}{Stock, P.}, \bibinfo{author}{Joulin, A.}, \bibinfo{author}{J{\'e}gou, H.}, \bibinfo{author}{Douze, M.}, \bibinfo{year}{2021}a.
\newblock \bibinfo{title}{Levit: a vision {Transformer} in convnet's clothing for faster inference}, in: \bibinfo{booktitle}{Proceedings of the IEEE International Conference on Computer Vision}, pp. \bibinfo{pages}{12259--12269}.
\newblock \DOIprefix\doi{10.1109/ICCV48922.2021.01204}.
\bibitem[{Graham et~al.(2021b)Graham, Jahanifar, Vu, Hadjigeorghiou, Leech, Snead, Raza, Minhas and Rajpoot}]{graham2021conic}
\bibinfo{author}{Graham, S.}, \bibinfo{author}{Jahanifar, M.}, \bibinfo{author}{Vu, Q.D.}, \bibinfo{author}{Hadjigeorghiou, G.}, \bibinfo{author}{Leech, T.}, \bibinfo{author}{Snead, D.}, \bibinfo{author}{Raza, S.E.A.}, \bibinfo{author}{Minhas, F.}, \bibinfo{author}{Rajpoot, N.}, \bibinfo{year}{2021}b.
\newblock \bibinfo{title}{{CoNIC}: Colon nuclei identification and counting challenge 2022}.
\newblock \bibinfo{journal}{arXiv:2111.14485} .
\bibitem[{Gu et~al.(2019)Gu, Cheng, Fu, Zhou, Hao, Zhao, Zhang, Gao and Liu}]{gu2019cenet}
\bibinfo{author}{Gu, Z.}, \bibinfo{author}{Cheng, J.}, \bibinfo{author}{Fu, H.}, \bibinfo{author}{Zhou, K.}, \bibinfo{author}{Hao, H.}, \bibinfo{author}{Zhao, Y.}, \bibinfo{author}{Zhang, T.}, \bibinfo{author}{Gao, S.}, \bibinfo{author}{Liu, J.}, \bibinfo{year}{2019}.
\newblock \bibinfo{title}{{CE-NET}: Context encoder network for {2D} medical image segmentation}.
\newblock \bibinfo{journal}{IEEE Transactions on Medical Imaging} \bibinfo{volume}{38}, \bibinfo{pages}{2281--2292}.
\newblock \DOIprefix\doi{10.1109/TMI.2019.2903562}.
\bibitem[{Gutman et~al.(2016)Gutman, Codella, Celebi, Helba, Marchetti, Mishra and Halpern}]{gutman2016skin}
\bibinfo{author}{Gutman, D.}, \bibinfo{author}{Codella, N.C.F.}, \bibinfo{author}{Celebi, E.}, \bibinfo{author}{Helba, B.}, \bibinfo{author}{Marchetti, M.}, \bibinfo{author}{Mishra, N.}, \bibinfo{author}{Halpern, A.}, \bibinfo{year}{2016}.
\newblock \bibinfo{title}{Skin lesion analysis toward melanoma detection: A challenge at the international symposium on biomedical imaging ({ISBI}) 2016, hosted by the international skin imaging collaboration ({ISIC})}, in: \bibinfo{booktitle}{arXiv:1605.01397}.
\bibitem[{Hatamizadeh et~al.(2024)Hatamizadeh, Heinrich, Yin, Tao, Alvarez, Kautz and Molchanov}]{hatamizadeh2023fastervit}
\bibinfo{author}{Hatamizadeh, A.}, \bibinfo{author}{Heinrich, G.}, \bibinfo{author}{Yin, H.}, \bibinfo{author}{Tao, A.}, \bibinfo{author}{Alvarez, J.M.}, \bibinfo{author}{Kautz, J.}, \bibinfo{author}{Molchanov, P.}, \bibinfo{year}{2024}.
\newblock \bibinfo{title}{{FasterViT}: Fast vision {Transformer}s with hierarchical attention}, in: \bibinfo{booktitle}{International Conference on Learning Representations}, pp. \bibinfo{pages}{1--14}.
\newblock \URLprefix \url{https://openreview.net/forum?id=kB4yBiNmXX}.
\bibitem[{Hatamizadeh et~al.(2022a)Hatamizadeh, Nath, Tang, Yang, Roth and Xu}]{hatamizadeh2022swin}
\bibinfo{author}{Hatamizadeh, A.}, \bibinfo{author}{Nath, V.}, \bibinfo{author}{Tang, Y.}, \bibinfo{author}{Yang, D.}, \bibinfo{author}{Roth, H.R.}, \bibinfo{author}{Xu, D.}, \bibinfo{year}{2022}a.
\newblock \bibinfo{title}{Swin {UNETR}: Swin {Transformer}s for semantic segmentation of brain tumors in {MRI} images}, in: \bibinfo{booktitle}{International Conference on Medical Image Computing and Computer Assisted Intervention Brainlesion Workshop}, pp. \bibinfo{pages}{272--284}.
\newblock \DOIprefix\doi{10.1007/978-3-031-08999-2_22}.
\bibitem[{Hatamizadeh et~al.(2022b)Hatamizadeh, Tang, Nath, Yang, Myronenko, Landman, Roth and Xu}]{hatamizadeh2022unetr}
\bibinfo{author}{Hatamizadeh, A.}, \bibinfo{author}{Tang, Y.}, \bibinfo{author}{Nath, V.}, \bibinfo{author}{Yang, D.}, \bibinfo{author}{Myronenko, A.}, \bibinfo{author}{Landman, B.}, \bibinfo{author}{Roth, H.R.}, \bibinfo{author}{Xu, D.}, \bibinfo{year}{2022}b.
\newblock \bibinfo{title}{{UNETR}: {Transformer}s for {3D} medical image segmentation}, in: \bibinfo{booktitle}{Proceedings of the IEEE/CVF Winter Conference on Applications of Computer Vision}, pp. \bibinfo{pages}{574--584}.
\newblock \DOIprefix\doi{10.1109/WACV51458.2022.00181}.
\bibitem[{Hatamizadeh et~al.(2019)Hatamizadeh, Terzopoulos and Myronenko}]{hatamizadeh2019end}
\bibinfo{author}{Hatamizadeh, A.}, \bibinfo{author}{Terzopoulos, D.}, \bibinfo{author}{Myronenko, A.}, \bibinfo{year}{2019}.
\newblock \bibinfo{title}{End-to-end boundary aware networks for medical image segmentation}, in: \bibinfo{booktitle}{International Workshop on Machine Learning in Medical Imaging}, pp. \bibinfo{pages}{187--194}.
\newblock \DOIprefix\doi{10.1007/978-3-030-32692-0_22}.
\bibitem[{He et~al.(2016)He, Zhang, Ren and Sun}]{he2016deep}
\bibinfo{author}{He, K.}, \bibinfo{author}{Zhang, X.}, \bibinfo{author}{Ren, S.}, \bibinfo{author}{Sun, J.}, \bibinfo{year}{2016}.
\newblock \bibinfo{title}{Deep residual learning for image recognition}, in: \bibinfo{booktitle}{Proceedings of the IEEE/CVF Conference on Computer Vision and Pattern Recognition}, pp. \bibinfo{pages}{770--778}.
\newblock \DOIprefix\doi{10.1109/CVPR.2016.90}.
\bibitem[{Huang et~al.(2021)Huang, Deng, Li and Yuan}]{huang2021missformer}
\bibinfo{author}{Huang, X.}, \bibinfo{author}{Deng, Z.}, \bibinfo{author}{Li, D.}, \bibinfo{author}{Yuan, X.}, \bibinfo{year}{2021}.
\newblock \bibinfo{title}{Missformer: An effective medical image segmentation {Transformer}}.
\newblock \bibinfo{journal}{IEEE Transactions on Medical Imaging} , \bibinfo{pages}{1484--1494}\DOIprefix\doi{10.1109/TMI.2022.3230943}.
\bibitem[{Irshad et~al.(2023)Irshad, Gomes and Kim}]{irshad2022improved}
\bibinfo{author}{Irshad, S.}, \bibinfo{author}{Gomes, D.P.}, \bibinfo{author}{Kim, S.T.}, \bibinfo{year}{2023}.
\newblock \bibinfo{title}{Improved abdominal multi-organ segmentation via 3{D} boundary-constrained deep neural networks}.
\newblock \bibinfo{journal}{IEEE Access} \bibinfo{volume}{11}, \bibinfo{pages}{35097--35110}.
\newblock \DOIprefix\doi{10.1109/ACCESS.2023.3264582}.
\bibitem[{Kamgar-Parsi and Rosenfeld(1999)}]{kamgar1999optimally}
\bibinfo{author}{Kamgar-Parsi, B.}, \bibinfo{author}{Rosenfeld, A.}, \bibinfo{year}{1999}.
\newblock \bibinfo{title}{Optimally isotropic laplacian operator}.
\newblock \bibinfo{journal}{IEEE Transactions on Image Processing} \bibinfo{volume}{8}, \bibinfo{pages}{1467--1472}.
\newblock \DOIprefix\doi{10.1109/83.791975}.
\bibitem[{Kanopoulos et~al.(1988)Kanopoulos, Vasanthavada and Baker}]{kanopoulos1988design}
\bibinfo{author}{Kanopoulos, N.}, \bibinfo{author}{Vasanthavada, N.}, \bibinfo{author}{Baker, R.L.}, \bibinfo{year}{1988}.
\newblock \bibinfo{title}{Design of an image edge detection filter using the {Sobel} operator}.
\newblock \bibinfo{journal}{IEEE Journal of Solid-State Circuits} \bibinfo{volume}{23}, \bibinfo{pages}{358--367}.
\newblock \DOIprefix\doi{10.1109/4.996}.
\bibitem[{Kervadec et~al.(2019)Kervadec, Bouchtiba, Desrosiers, Granger, Dolz and Ayed}]{kervadec2019boundary}
\bibinfo{author}{Kervadec, H.}, \bibinfo{author}{Bouchtiba, J.}, \bibinfo{author}{Desrosiers, C.}, \bibinfo{author}{Granger, E.}, \bibinfo{author}{Dolz, J.}, \bibinfo{author}{Ayed, I.B.}, \bibinfo{year}{2019}.
\newblock \bibinfo{title}{Boundary loss for highly unbalanced segmentation}, in: \bibinfo{booktitle}{International Conference on Medical Imaging with Deep Learning}, \bibinfo{organization}{PMLR}. pp. \bibinfo{pages}{285--296}.
\newblock \URLprefix \url{https://proceedings.mlr.press/v102/kervadec19a.html}.
\bibitem[{Lee et~al.(2020)Lee, Kim, Lee, Kim and Ro}]{lee2020structure}
\bibinfo{author}{Lee, H.J.}, \bibinfo{author}{Kim, J.U.}, \bibinfo{author}{Lee, S.}, \bibinfo{author}{Kim, H.G.}, \bibinfo{author}{Ro, Y.M.}, \bibinfo{year}{2020}.
\newblock \bibinfo{title}{Structure boundary preserving segmentation for medical image with ambiguous boundary}, in: \bibinfo{booktitle}{Proceedings of the IEEE/CVF Conference on Computer Vision and Pattern Recognition}, pp. \bibinfo{pages}{4817--4826}.
\newblock \DOIprefix\doi{10.1109/CVPR42600.2020.00487}.
\bibitem[{Li et~al.(2018)Li, Zia, Tran, Yu, Hager and Chandraker}]{li2018deep}
\bibinfo{author}{Li, C.}, \bibinfo{author}{Zia, M.Z.}, \bibinfo{author}{Tran, Q.H.}, \bibinfo{author}{Yu, X.}, \bibinfo{author}{Hager, G.D.}, \bibinfo{author}{Chandraker, M.}, \bibinfo{year}{2018}.
\newblock \bibinfo{title}{Deep supervision with intermediate concepts}.
\newblock \bibinfo{journal}{IEEE Transactions on Pattern Analysis and Machine Intelligence} \bibinfo{volume}{41}, \bibinfo{pages}{1828--1843}.
\newblock \DOIprefix\doi{10.1109/TPAMI.2018.2863285}.
\bibitem[{Lin et~al.(2025)Lin, Fang, Zhang, Cheng and Chen}]{lin2023permutable}
\bibinfo{author}{Lin, Y.}, \bibinfo{author}{Fang, X.}, \bibinfo{author}{Zhang, D.}, \bibinfo{author}{Cheng, K.T.}, \bibinfo{author}{Chen, H.}, \bibinfo{year}{2025}.
\newblock \bibinfo{title}{Boosting convolution with efficient {MLP}-permutation for volumetric medical image segmentation}.
\newblock \bibinfo{journal}{IEEE Transactions on Medical Imaging} , \bibinfo{pages}{1--1}\DOIprefix\doi{10.1109/TMI.2025.3530113}.
\bibitem[{Lin et~al.(2021)Lin, Liu, Ma and Zheng}]{lin2021seg4reg}
\bibinfo{author}{Lin, Y.}, \bibinfo{author}{Liu, L.}, \bibinfo{author}{Ma, K.}, \bibinfo{author}{Zheng, Y.}, \bibinfo{year}{2021}.
\newblock \bibinfo{title}{Seg4reg+: Consistency learning between spine segmentation and cobb angle regression}, in: \bibinfo{booktitle}{International Conference on Medical Image Computing and Computer Assisted Intervention}, \bibinfo{publisher}{Lecture Notes in Computer Science, vol 12905. Springer, Cham}. pp. \bibinfo{pages}{490--499}.
\newblock \DOIprefix\doi{10.1007/978-3-030-87240-3_47}.
\bibitem[{Lin et~al.(2024)Lin, Liu, Chen, Yang, Ma, Zheng and Cheng}]{10528362}
\bibinfo{author}{Lin, Y.}, \bibinfo{author}{Liu, Y.}, \bibinfo{author}{Chen, H.}, \bibinfo{author}{Yang, X.}, \bibinfo{author}{Ma, K.}, \bibinfo{author}{Zheng, Y.}, \bibinfo{author}{Cheng, K.T.}, \bibinfo{year}{2024}.
\newblock \bibinfo{title}{{LENAS}: Learning-based neural architecture search and ensemble for 3-{D} radiotherapy dose prediction}.
\newblock \bibinfo{journal}{IEEE Transactions on Cybernetics} \bibinfo{volume}{54}, \bibinfo{pages}{5795--5805}.
\newblock \DOIprefix\doi{10.1109/tcyb.2024.3390769}.
\bibitem[{Lin et~al.(2023a)Lin, Qu, Chen, Gao, Li, Xia, Ma, Zheng and Cheng}]{lin2022label}
\bibinfo{author}{Lin, Y.}, \bibinfo{author}{Qu, Z.}, \bibinfo{author}{Chen, H.}, \bibinfo{author}{Gao, Z.}, \bibinfo{author}{Li, Y.}, \bibinfo{author}{Xia, L.}, \bibinfo{author}{Ma, K.}, \bibinfo{author}{Zheng, Y.}, \bibinfo{author}{Cheng, K.T.}, \bibinfo{year}{2023}a.
\newblock \bibinfo{title}{Nuclei segmentation with point annotations from pathology images via self-supervised learning and co-training}.
\newblock \bibinfo{journal}{Medical Image Analysis} \bibinfo{volume}{89}, \bibinfo{pages}{102933}.
\newblock \DOIprefix\doi{10.1016/j.media.2023.102933}.
\bibitem[{Lin et~al.(2019)Lin, Su, Wang, Li, Liu, Cheng and Yang}]{lin2019automated}
\bibinfo{author}{Lin, Y.}, \bibinfo{author}{Su, J.}, \bibinfo{author}{Wang, X.}, \bibinfo{author}{Li, X.}, \bibinfo{author}{Liu, J.}, \bibinfo{author}{Cheng, K.T.}, \bibinfo{author}{Yang, X.}, \bibinfo{year}{2019}.
\newblock \bibinfo{title}{Automated pulmonary embolism detection from {CTPA} images using an end-to-end convolutional neural network}, in: \bibinfo{booktitle}{International Conference on Medical Image Computing and Computer Assisted Intervention}, \bibinfo{publisher}{Lecture Notes in Computer Science, vol 11767. Springer, Cham}. pp. \bibinfo{pages}{280--288}.
\newblock \DOIprefix\doi{10.1007/978-3-030-32251-9_31}.
\bibitem[{Lin et~al.(2023b)Lin, Zhang, Fang, Chen, Cheng and Chen}]{lin2023rethinking}
\bibinfo{author}{Lin, Y.}, \bibinfo{author}{Zhang, D.}, \bibinfo{author}{Fang, X.}, \bibinfo{author}{Chen, Y.}, \bibinfo{author}{Cheng, K.T.}, \bibinfo{author}{Chen, H.}, \bibinfo{year}{2023}b.
\newblock \bibinfo{title}{Rethinking boundary detection in deep learning models for medical image segmentation}, in: \bibinfo{booktitle}{International Conference on Information Processing in Medical Imaging}, \bibinfo{publisher}{Lecture Notes in Computer Science, vol 13939. Springer, Cham}. pp. \bibinfo{pages}{730--742}.
\newblock \DOIprefix\doi{10.1007/978-3-031-34048-2_56}.
\bibitem[{Liu et~al.(2022)Liu, Hu, Lin, Yao, Xie, Wei, Ning, Cao, Zhang, Dong et~al.}]{liu2022swin}
\bibinfo{author}{Liu, Z.}, \bibinfo{author}{Hu, H.}, \bibinfo{author}{Lin, Y.}, \bibinfo{author}{Yao, Z.}, \bibinfo{author}{Xie, Z.}, \bibinfo{author}{Wei, Y.}, \bibinfo{author}{Ning, J.}, \bibinfo{author}{Cao, Y.}, \bibinfo{author}{Zhang, Z.}, \bibinfo{author}{Dong, L.}, et~al., \bibinfo{year}{2022}.
\newblock \bibinfo{title}{Swin {{Transformer}} v2: Scaling up capacity and resolution}, in: \bibinfo{booktitle}{Proceedings of the IEEE/CVF Conference on Computer Vision and Pattern Recognition}, pp. \bibinfo{pages}{12009--12019}.
\newblock \DOIprefix\doi{10.1109/CVPR52688.2022.01170}.
\bibitem[{Long et~al.(2015)Long, Shelhamer and Darrell}]{long2015fully}
\bibinfo{author}{Long, J.}, \bibinfo{author}{Shelhamer, E.}, \bibinfo{author}{Darrell, T.}, \bibinfo{year}{2015}.
\newblock \bibinfo{title}{Fully convolutional networks for semantic segmentation}, in: \bibinfo{booktitle}{Proceedings of the IEEE/CVF Conference on Computer Vision and Pattern Recognition}, pp. \bibinfo{pages}{3431--3440}.
\newblock \DOIprefix\doi{10.1109/CVPR.2015.7298965}.
\bibitem[{Ma et~al.(2024)Ma, He, Li, Han, You and Wang}]{ma2023segment}
\bibinfo{author}{Ma, J.}, \bibinfo{author}{He, Y.}, \bibinfo{author}{Li, F.}, \bibinfo{author}{Han, L.}, \bibinfo{author}{You, C.}, \bibinfo{author}{Wang, B.}, \bibinfo{year}{2024}.
\newblock \bibinfo{title}{Segment anything in medical images}.
\newblock \bibinfo{journal}{Nature Communications} \bibinfo{volume}{15}, \bibinfo{pages}{654}.
\newblock \DOIprefix\doi{10.1038/s41467-024-44824-z}.
\bibitem[{Mendon{\c{c}}a et~al.(2013)Mendon{\c{c}}a, Ferreira, Marques, Marcal and Rozeira}]{mendoncca2013ph}
\bibinfo{author}{Mendon{\c{c}}a, T.}, \bibinfo{author}{Ferreira, P.M.}, \bibinfo{author}{Marques, J.S.}, \bibinfo{author}{Marcal, A.R.}, \bibinfo{author}{Rozeira, J.}, \bibinfo{year}{2013}.
\newblock \bibinfo{title}{{PH 2-A} dermoscopic image database for research and benchmarking}, in: \bibinfo{booktitle}{International Conference of the IEEE Engineering in Medicine and Biology Society}, pp. \bibinfo{pages}{5437--5440}.
\newblock \DOIprefix\doi{10.1109/EMBC.2013.6610779}.
\bibitem[{Milletari et~al.(2016)Milletari, Navab and Ahmadi}]{milletari2016v}
\bibinfo{author}{Milletari, F.}, \bibinfo{author}{Navab, N.}, \bibinfo{author}{Ahmadi, S.A.}, \bibinfo{year}{2016}.
\newblock \bibinfo{title}{V-{Net}: Fully convolutional neural networks for volumetric medical image segmentation}, in: \bibinfo{booktitle}{International Conference on 3D Vision}, pp. \bibinfo{pages}{565--571}.
\newblock \DOIprefix\doi{10.1109/3DV.2016.79}.
\bibitem[{Peng et~al.(2005)Peng, Long and Ding}]{peng2005feature}
\bibinfo{author}{Peng, H.}, \bibinfo{author}{Long, F.}, \bibinfo{author}{Ding, C.}, \bibinfo{year}{2005}.
\newblock \bibinfo{title}{Feature selection based on mutual information criteria of max-dependency, max-relevance, and min-redundancy}.
\newblock \bibinfo{journal}{IEEE Transactions on Pattern Analysis and Machine Intelligence} \bibinfo{volume}{27}, \bibinfo{pages}{1226--1238}.
\newblock \DOIprefix\doi{10.1109/TPAMI.2005.159}.
\bibitem[{Ronneberger et~al.(2015)Ronneberger, Fischer and Brox}]{ronneberger2015u}
\bibinfo{author}{Ronneberger, O.}, \bibinfo{author}{Fischer, P.}, \bibinfo{author}{Brox, T.}, \bibinfo{year}{2015}.
\newblock \bibinfo{title}{U-{Net}: Convolutional networks for biomedical image segmentation}, in: \bibinfo{booktitle}{International Conference on Medical Image Computing and Computer Assisted Intervention}, \bibinfo{publisher}{Lecture Notes in Computer Science, vol 9351. Springer, Cham}. pp. \bibinfo{pages}{234--241}.
\newblock \DOIprefix\doi{10.1007/978-3-319-24574-4_28}.
\bibitem[{Schlemper et~al.(2019)Schlemper, Oktay, Schaap, Heinrich, Kainz, Glocker and Rueckert}]{schlemper2019attention}
\bibinfo{author}{Schlemper, J.}, \bibinfo{author}{Oktay, O.}, \bibinfo{author}{Schaap, M.}, \bibinfo{author}{Heinrich, M.}, \bibinfo{author}{Kainz, B.}, \bibinfo{author}{Glocker, B.}, \bibinfo{author}{Rueckert, D.}, \bibinfo{year}{2019}.
\newblock \bibinfo{title}{Attention gated networks: Learning to leverage salient regions in medical images}.
\newblock \bibinfo{journal}{Medical Image Analysis} \bibinfo{volume}{53}, \bibinfo{pages}{197--207}.
\newblock \DOIprefix\doi{10.1016/j.media.2019.01.012}.
\bibitem[{Shaker et~al.(2024)Shaker, Maaz, Rasheed, Khan, Yang and Khan}]{shaker2024unetr}
\bibinfo{author}{Shaker, A.M.}, \bibinfo{author}{Maaz, M.}, \bibinfo{author}{Rasheed, H.}, \bibinfo{author}{Khan, S.}, \bibinfo{author}{Yang, M.H.}, \bibinfo{author}{Khan, F.S.}, \bibinfo{year}{2024}.
\newblock \bibinfo{title}{{UNETR}++: delving into efficient and accurate {3D} medical image segmentation}.
\newblock \bibinfo{journal}{IEEE Transactions on Medical Imaging} , \bibinfo{pages}{3377--3390}\DOIprefix\doi{10.1109/TMI.2024.3398728}.
\bibitem[{Shamshad et~al.(2023)Shamshad, Khan, Zamir, Khan, Hayat, Khan and Fu}]{shamshad2022transformers}
\bibinfo{author}{Shamshad, F.}, \bibinfo{author}{Khan, S.}, \bibinfo{author}{Zamir, S.W.}, \bibinfo{author}{Khan, M.H.}, \bibinfo{author}{Hayat, M.}, \bibinfo{author}{Khan, F.S.}, \bibinfo{author}{Fu, H.}, \bibinfo{year}{2023}.
\newblock \bibinfo{title}{Transformers in medical imaging: A survey}.
\newblock \bibinfo{journal}{Medical Image Analysis} \bibinfo{volume}{88}, \bibinfo{pages}{102802}.
\newblock \DOIprefix\doi{10.1016/j.media.2023.102802}.
\bibitem[{Shi et~al.(2023)Shi, Chen and Zhang}]{shi2023transformer}
\bibinfo{author}{Shi, Z.}, \bibinfo{author}{Chen, H.}, \bibinfo{author}{Zhang, D.}, \bibinfo{year}{2023}.
\newblock \bibinfo{title}{Transformer-auxiliary neural networks for image manipulation localization by operator inductions}.
\newblock \bibinfo{journal}{IEEE Transactions on Circuits and Systems for Video Technology} , \bibinfo{pages}{4907--4920}\DOIprefix\doi{10.1109/TCSVT.2023.32514}.
\bibitem[{Sun et~al.(2023)Sun, Luo and Li}]{sun2023boundary}
\bibinfo{author}{Sun, F.}, \bibinfo{author}{Luo, Z.}, \bibinfo{author}{Li, S.}, \bibinfo{year}{2023}.
\newblock \bibinfo{title}{Boundary difference over union loss for medical image segmentation}, in: \bibinfo{booktitle}{International Conference on Medical Image Computing and Computer Assisted Intervention}, \bibinfo{publisher}{Lecture Notes in Computer Science, vol 14223. Springer, Cham}. pp. \bibinfo{pages}{292--301}.
\newblock \DOIprefix\doi{10.1007/978-3-031-43901-8_28}.
\bibitem[{Valanarasu et~al.(2021)Valanarasu, Oza, Hacihaliloglu and Patel}]{valanarasu2021medical}
\bibinfo{author}{Valanarasu, J.M.J.}, \bibinfo{author}{Oza, P.}, \bibinfo{author}{Hacihaliloglu, I.}, \bibinfo{author}{Patel, V.M.}, \bibinfo{year}{2021}.
\newblock \bibinfo{title}{Medical {Transformer}: Gated axial-attention for medical image segmentation}, in: \bibinfo{booktitle}{International Conference on Medical Image Computing and Computer Assisted Intervention}, \bibinfo{organization}{Lecture Notes in Computer Science, vol 12901. Springer, Cham}. pp. \bibinfo{pages}{36--46}.
\newblock \DOIprefix\doi{10.1007/978-3-030-87193-2_4}.
\bibitem[{Vaswani et~al.(2017)Vaswani, Shazeer, Parmar, Uszkoreit, Jones, Gomez, Kaiser and Polosukhin}]{vaswani2017attention}
\bibinfo{author}{Vaswani, A.}, \bibinfo{author}{Shazeer, N.}, \bibinfo{author}{Parmar, N.}, \bibinfo{author}{Uszkoreit, J.}, \bibinfo{author}{Jones, L.}, \bibinfo{author}{Gomez, A.N.}, \bibinfo{author}{Kaiser, {\L}.}, \bibinfo{author}{Polosukhin, I.}, \bibinfo{year}{2017}.
\newblock \bibinfo{title}{Attention is all you need}, in: \bibinfo{booktitle}{Advances in Neural Information Processing Systems}, \bibinfo{publisher}{Curran Associates, Inc.}. pp. \bibinfo{pages}{5998--6008}.
\newblock \URLprefix \url{http://arxiv.org/abs/1706.03762}.
\bibitem[{Wang et~al.(2023)Wang, Chen, Ma, Wang, Fei, Shuai, Tang, Zhou and Qin}]{wang2023xbound}
\bibinfo{author}{Wang, J.}, \bibinfo{author}{Chen, F.}, \bibinfo{author}{Ma, Y.}, \bibinfo{author}{Wang, L.}, \bibinfo{author}{Fei, Z.}, \bibinfo{author}{Shuai, J.}, \bibinfo{author}{Tang, X.}, \bibinfo{author}{Zhou, Q.}, \bibinfo{author}{Qin, J.}, \bibinfo{year}{2023}.
\newblock \bibinfo{title}{Xbound-former: Toward cross-scale boundary modeling in transformers}.
\newblock \bibinfo{journal}{IEEE Transactions on Medical Imaging} \bibinfo{volume}{42}, \bibinfo{pages}{1735--1745}.
\newblock \DOIprefix\doi{10.1109/TMI.2023.3236037}.
\bibitem[{Wang et~al.(2021)Wang, Wei, Wang, Zhou, Zhu and Qin}]{wang2021boundary}
\bibinfo{author}{Wang, J.}, \bibinfo{author}{Wei, L.}, \bibinfo{author}{Wang, L.}, \bibinfo{author}{Zhou, Q.}, \bibinfo{author}{Zhu, L.}, \bibinfo{author}{Qin, J.}, \bibinfo{year}{2021}.
\newblock \bibinfo{title}{Boundary-aware {Transformer}s for skin lesion segmentation}, in: \bibinfo{booktitle}{International Conference on Medical Image Computing and Computer Assisted Intervention}, \bibinfo{publisher}{Lecture Notes in Computer Science, vol 12901. Springer, Cham}. pp. \bibinfo{pages}{206--216}.
\newblock \DOIprefix\doi{10.1007/978-3-030-87193-2_20}.
\bibitem[{Wang et~al.(2022)Wang, Chen, Ji, Fan and Li}]{wang2022boundary}
\bibinfo{author}{Wang, R.}, \bibinfo{author}{Chen, S.}, \bibinfo{author}{Ji, C.}, \bibinfo{author}{Fan, J.}, \bibinfo{author}{Li, Y.}, \bibinfo{year}{2022}.
\newblock \bibinfo{title}{Boundary-aware context neural network for medical image segmentation}.
\newblock \bibinfo{journal}{Medical Image Analysis} \bibinfo{volume}{78}, \bibinfo{pages}{102395}.
\newblock \DOIprefix\doi{10.1016/j.media.2022.102395}.
\bibitem[{Wang(2007)}]{wang2007laplacian}
\bibinfo{author}{Wang, X.}, \bibinfo{year}{2007}.
\newblock \bibinfo{title}{Laplacian operator-based edge detectors}.
\newblock \bibinfo{journal}{IEEE Transactions on Pattern Analysis and Machine Intelligence} \bibinfo{volume}{29}, \bibinfo{pages}{886--890}.
\newblock \DOIprefix\doi{10.1109/TPAMI.2007.1027}.
\bibitem[{Wijeratne et~al.(2021)Wijeratne, Alexander, Initiative et~al.}]{wijeratne2021learning}
\bibinfo{author}{Wijeratne, P.A.}, \bibinfo{author}{Alexander, D.C.}, \bibinfo{author}{Initiative, A.D.N.}, et~al., \bibinfo{year}{2021}.
\newblock \bibinfo{title}{Learning transition times in event sequences: The temporal event-based model of disease progression}, in: \bibinfo{booktitle}{International Conference on Information Processing in Medical Imaging}, \bibinfo{publisher}{Lecture Notes in Computer Science, vol 12729. Springer, Cham}. pp. \bibinfo{pages}{583--595}.
\newblock \DOIprefix\doi{10.1007/978-3-030-78191-0_45}.
\bibitem[{Wu et~al.(2022)Wu, Fang, Shang, Yang, Wang, Gao, Yang and Xu}]{wu2022seatrans}
\bibinfo{author}{Wu, J.}, \bibinfo{author}{Fang, H.}, \bibinfo{author}{Shang, F.}, \bibinfo{author}{Yang, D.}, \bibinfo{author}{Wang, Z.}, \bibinfo{author}{Gao, J.}, \bibinfo{author}{Yang, Y.}, \bibinfo{author}{Xu, Y.}, \bibinfo{year}{2022}.
\newblock \bibinfo{title}{{SeATrans}: Learning segmentation-assisted diagnosis model via {Transformer}}, in: \bibinfo{booktitle}{International Conference on Medical Image Computing and Computer Assisted Intervention}, \bibinfo{organization}{Lecture Notes in Computer Science, vol 13432. Springer, Cham}. pp. \bibinfo{pages}{677--687}.
\newblock \DOIprefix\doi{10.1007/978-3-031-16434-7_65}.
\bibitem[{Xiao et~al.(2023)Xiao, Li, Liu, Zhu and Zhang}]{xiao2023transformers}
\bibinfo{author}{Xiao, H.}, \bibinfo{author}{Li, L.}, \bibinfo{author}{Liu, Q.}, \bibinfo{author}{Zhu, X.}, \bibinfo{author}{Zhang, Q.}, \bibinfo{year}{2023}.
\newblock \bibinfo{title}{Transformers in medical image segmentation: A review}.
\newblock \bibinfo{journal}{Biomedical Signal Processing and Control} \bibinfo{volume}{84}, \bibinfo{pages}{104791}.
\newblock \DOIprefix\doi{https://doi.org/10.1016/j.bspc.2023.104791}.
\bibitem[{Xie et~al.(2017)Xie, Girshick, Doll{\'a}r, Tu and He}]{xie2017aggregated}
\bibinfo{author}{Xie, S.}, \bibinfo{author}{Girshick, R.}, \bibinfo{author}{Doll{\'a}r, P.}, \bibinfo{author}{Tu, Z.}, \bibinfo{author}{He, K.}, \bibinfo{year}{2017}.
\newblock \bibinfo{title}{Aggregated residual transformations for deep neural networks}, in: \bibinfo{booktitle}{Proceedings of the IEEE/CVF Conference on Computer Vision and Pattern Recognition}, pp. \bibinfo{pages}{1492--1500}.
\newblock \DOIprefix\doi{10.1109/CVPR.2017.634}.
\bibitem[{Yan et~al.(2022)Yan, Tang, Sun, Ma, Kong and Xie}]{yan2022after}
\bibinfo{author}{Yan, X.}, \bibinfo{author}{Tang, H.}, \bibinfo{author}{Sun, S.}, \bibinfo{author}{Ma, H.}, \bibinfo{author}{Kong, D.}, \bibinfo{author}{Xie, X.}, \bibinfo{year}{2022}.
\newblock \bibinfo{title}{{After-UNet}: Axial fusion {Transformer} {U}net for medical image segmentation}, in: \bibinfo{booktitle}{Proceedings of the IEEE/CVF Winter Conference on Applications of Computer Vision}, pp. \bibinfo{pages}{3971--3981}.
\newblock \DOIprefix\doi{10.1109/WACV51458.2022.00333}.
\bibitem[{Yu and Helwig(2022)}]{yu2022role}
\bibinfo{author}{Yu, C.}, \bibinfo{author}{Helwig, E.J.}, \bibinfo{year}{2022}.
\newblock \bibinfo{title}{The role of {AI} technology in prediction, diagnosis and treatment of colorectal cancer}.
\newblock \bibinfo{journal}{Artificial Intelligence Review} , \bibinfo{pages}{323--343}\DOIprefix\doi{10.1007/s10462-021-10034-y}.
\bibitem[{Yuan et~al.(2023)Yuan, Zhang and Fang}]{yuan2023effective}
\bibinfo{author}{Yuan, F.}, \bibinfo{author}{Zhang, Z.}, \bibinfo{author}{Fang, Z.}, \bibinfo{year}{2023}.
\newblock \bibinfo{title}{An effective {CNN} and {Transformer} complementary network for medical image segmentation}.
\newblock \bibinfo{journal}{Pattern Recognition} \bibinfo{volume}{136}, \bibinfo{pages}{109228}.
\newblock \DOIprefix\doi{10.1016/j.patcog.2022.109228}.
\bibitem[{Zhang et~al.(2022a)Zhang, Lin, Chen, Tian, Yang, Tang and Cheng}]{zhang2022deep}
\bibinfo{author}{Zhang, D.}, \bibinfo{author}{Lin, Y.}, \bibinfo{author}{Chen, H.}, \bibinfo{author}{Tian, Z.}, \bibinfo{author}{Yang, X.}, \bibinfo{author}{Tang, J.}, \bibinfo{author}{Cheng, K.T.}, \bibinfo{year}{2022}a.
\newblock \bibinfo{title}{Understanding the tricks of deep learning in medical image segmentation: Challenges and future directions}.
\newblock \bibinfo{journal}{arXiv preprint arXiv:2209.10307} .
\bibitem[{Zhang et~al.(2022b)Zhang, Tang and Cheng}]{zhang2022graph}
\bibinfo{author}{Zhang, D.}, \bibinfo{author}{Tang, J.}, \bibinfo{author}{Cheng, K.T.}, \bibinfo{year}{2022}b.
\newblock \bibinfo{title}{Graph reasoning {Transformer} for image parsing}, in: \bibinfo{booktitle}{International Conference on Multimedia}, pp. \bibinfo{pages}{2380--2389}.
\newblock \DOIprefix\doi{10.1145/3503161.354785}.
\bibitem[{Zhang et~al.(2020a)Zhang, Zhang, Tang, Hua and Sun}]{NEURIPS2020_07211688}
\bibinfo{author}{Zhang, D.}, \bibinfo{author}{Zhang, H.}, \bibinfo{author}{Tang, J.}, \bibinfo{author}{Hua, X.S.}, \bibinfo{author}{Sun, Q.}, \bibinfo{year}{2020}a.
\newblock \bibinfo{title}{Causal intervention for weakly-supervised semantic segmentation}, in: \bibinfo{booktitle}{Advances in Neural Information Processing Systems}, pp. \bibinfo{pages}{655--666}.
\bibitem[{Zhang et~al.(2021a)Zhang, Zhang, Tang, Hua and Sun}]{zhang2021self}
\bibinfo{author}{Zhang, D.}, \bibinfo{author}{Zhang, H.}, \bibinfo{author}{Tang, J.}, \bibinfo{author}{Hua, X.S.}, \bibinfo{author}{Sun, Q.}, \bibinfo{year}{2021}a.
\newblock \bibinfo{title}{Self-regulation for semantic segmentation}, in: \bibinfo{booktitle}{Proceedings of the IEEE International Conference on Computer Vision}, pp. \bibinfo{pages}{6953--6963}.
\newblock \DOIprefix\doi{10.1109/ICCV48922.2021.00687}.
\bibitem[{Zhang et~al.(2020b)Zhang, Zhang, Tang, Wang, Hua and Sun}]{zhang2020feature}
\bibinfo{author}{Zhang, D.}, \bibinfo{author}{Zhang, H.}, \bibinfo{author}{Tang, J.}, \bibinfo{author}{Wang, M.}, \bibinfo{author}{Hua, X.}, \bibinfo{author}{Sun, Q.}, \bibinfo{year}{2020}b.
\newblock \bibinfo{title}{Feature pyramid transformer}, in: \bibinfo{booktitle}{Computer Vision--ECCV 2020: 16th European Conference, Glasgow, UK, August 23--28, 2020, Proceedings, Part XXVIII 16}, \bibinfo{organization}{Springer}. pp. \bibinfo{pages}{323--339}.
\newblock \DOIprefix\doi{10.1007/978-3-030-58604-1_20}.
\bibitem[{Zhang et~al.(2023)Zhang, Zhang and Tang}]{zhang2023augmented}
\bibinfo{author}{Zhang, D.}, \bibinfo{author}{Zhang, L.}, \bibinfo{author}{Tang, J.}, \bibinfo{year}{2023}.
\newblock \bibinfo{title}{Augmented {FCN}: rethinking context modeling for semantic segmentation}.
\newblock \bibinfo{journal}{Science China Information Sciences} \bibinfo{volume}{66}, \bibinfo{pages}{142105}.
\newblock \DOIprefix\doi{10.1007/s11432-021-3590-1}.
\bibitem[{Zhang et~al.(2021b)Zhang, Liu and Hu}]{zhang2021transfuse}
\bibinfo{author}{Zhang, Y.}, \bibinfo{author}{Liu, H.}, \bibinfo{author}{Hu, Q.}, \bibinfo{year}{2021}b.
\newblock \bibinfo{title}{Transfuse: Fusing {Transformer}s and {CNN}s for medical image segmentation}, in: \bibinfo{booktitle}{International Conference on Medical Image Computing and Computer Assisted Intervention}, \bibinfo{organization}{Lecture Notes in Computer Science, vol 12901. Springer, Cham}. pp. \bibinfo{pages}{14--24}.
\newblock \DOIprefix\doi{10.1007/978-3-030-87193-2_2}.
\bibitem[{Zhao et~al.(2024)Zhao, Zhong and Wang}]{zhao2024semi}
\bibinfo{author}{Zhao, W.}, \bibinfo{author}{Zhong, L.}, \bibinfo{author}{Wang, G.}, \bibinfo{year}{2024}.
\newblock \bibinfo{title}{{SEMI-CONTRANS}: Semi-supervised medical image segmentation via multi-scale feature fusion and cross teaching of {CNN} and {Transformer}}, in: \bibinfo{booktitle}{IEEE International Symposium on Biomedical Imaging}, \bibinfo{organization}{IEEE}. pp. \bibinfo{pages}{1--5}.
\newblock \DOIprefix\doi{10.1109/ISBI56570.2024.10635274}.
\bibitem[{Zhou et~al.(2018)Zhou, Siddiquee, Tajbakhsh and Liang}]{zhou2018unetpp}
\bibinfo{author}{Zhou, Z.}, \bibinfo{author}{Siddiquee, M.M.R.}, \bibinfo{author}{Tajbakhsh, N.}, \bibinfo{author}{Liang, J.}, \bibinfo{year}{2018}.
\newblock \bibinfo{title}{{UNet}++: A nested {U-Net} architecture for medical image segmentation}.
\newblock \bibinfo{journal}{Deep Learning in Medical Image Analysis and Multimodal Learning for Clinical Decision Suppor} , \bibinfo{pages}{3--11}\DOIprefix\doi{10.1007/978-3-030-00889-5_1}.

\end{thebibliography}
\end{document}